\documentclass[journal=esthag,manuscript=article,layout=twocolumn]{achemso}

\usepackage[utf8]{inputenc}
\usepackage[T1]{fontenc}
\usepackage[final]{microtype}
\usepackage[american]{babel}
\usepackage{graphicx}
\usepackage{booktabs}

\usepackage{threeparttable}
\usepackage[version=4]{mhchem}
\usepackage{siunitx}
\usepackage{amsmath}
\usepackage{makecell}

\DeclareSIUnit\angstrom{\text{Å}}
\DeclareSIUnit\kcal{\text{kcal}}

\author{Rinto Thomas}
\affiliation[UMR]{Fachbereich Chemie, Philipps-Universität Marburg, 35032 Marburg, Germany}
\altaffiliation{R.T. and P.R.P. contributed equally to this work.}
\author{Praveen Ranganath Prabhakar}
\affiliation[UCI]{Department of Chemistry, University of California, Irvine, Irvine, California, 92697 United States}
\altaffiliation{R.T. and P.R.P. contributed equally to this work.}
\author{Douglas J. Tobias}
\affiliation[UCI]{Department of Chemistry, University of California, Irvine, Irvine, California, 92697 United States}
\author{Michael von Domaros}
\affiliation[UMR]{Fachbereich Chemie, Philipps-Universität Marburg, 35032 Marburg, Germany}
\email{mvondomaros@uni-marburg.de}

\title{Modeling Diffusion and Permeation Across the Stratum Corneum Lipid Barrier}

\begin{document}

\begin{tocentry}
    \includegraphics[width=3.25in]{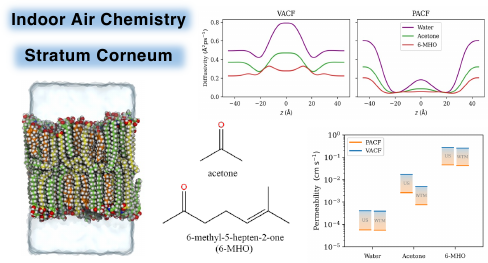}
\end{tocentry}

\begin{abstract}
Human skin oils are a major sink for ozone in densely occupied indoor environments. 
Understanding how the resulting volatile and semivolatile organic oxidation products influence indoor air chemistry requires accurate representations not only of their emission into indoor air but also of their transport across the outermost skin barrier, the stratum corneum. 
Using molecular dynamics simulations, we investigate the passive permeation of acetone, 6-methyl-5-hepten-2-one, and water—two representative products of skin-oil oxidation and a reference compound—through a model stratum corneum lipid membrane. 
We determine position-dependent diffusivities using two complementary analyses based on the same set of simulations and evaluate their accuracy through a propagator analysis. 
The two approaches provide upper and lower bounds for the true diffusivity, which, when combined with previously reported free-energy profiles, yield permeabilities relevant for modeling macroscopic skin transport. 
Our results show that permeation is governed primarily by energetic barriers rather than by molecular mobility, and that the predicted transport coefficients vary by about one order of magnitude depending on the chosen diffusivity estimator. 
These findings provide molecular-level constraints for parameters used in indoor air chemistry models and establish a transferable framework for linking atomistic transport mechanisms to large-scale simulations of human exposure and indoor air quality.
\end{abstract}

\section{Introduction}

In 2024, the mean global near-surface temperature exceeded the pre-industrial average by more than \SI{1.5}{K},\cite{WorldMeteorologicalOrganizationWMO.WorldMeteorologicalOrganizationWMO.2025.StateGlobalClimate} underscoring the increasingly tangible impacts of global warming on human health, behavior, and daily life.\cite{Rocque.Witteman.2021.HealthEffectsClimate,IntergovernmentalPanelOnClimateChangeIPCC.IntergovernmentalPanelOnClimateChangeIPCC.2022.Impacts15degCGlobal,Abdul-Nabi.Zahran.2025.ClimateChangeIts}
A plausible behavioral response—particularly among vulnerable populations such as older adults and individuals with pre-existing health conditions—is to spend more time indoors to mitigate thermal discomfort during periods of extreme heat. However, empirical evidence remains limited regarding how climate change is reshaping patterns of time use.\cite{White-Newsome.ONeill.2012.ClimateChangeHealth,Connolly.Connolly.2018.ClimateChangeAllocation}

Nevertheless, because humans spend most of their time indoors, the impacts of climate change are largely mediated through indoor environments.\cite{InstituteOfMedicine.InstituteOfMedicine.2011.ClimateChangeIndoor,Spengler.Spengler.2012.ClimateChangeIndoor}
Accordingly, understanding how rising outdoor temperatures influence indoor air quality has become an increasingly important research priority.\cite{Nazaroff.Nazaroff.2013.ExploringConsequencesClimate,Mansouri.Blondeau.2022.ImpactClimateChange}
Models and simulations are expected to play a central role in evaluating the long-term effects of climate change on indoor environments.\cite{Zhao.Schieweck.2024.LongtermPredictionEffects,Zhao.Hussein.2025.LongtermPredictionClimate}
However, these modeling efforts rely on accurate physicochemical representations of the complex surface–air chemistry occurring indoors.\cite{VonDomaros.Tobias.2025.MolecularDynamicsSimulations}

\begin{figure}
    \centering
    \includegraphics[width=\columnwidth]{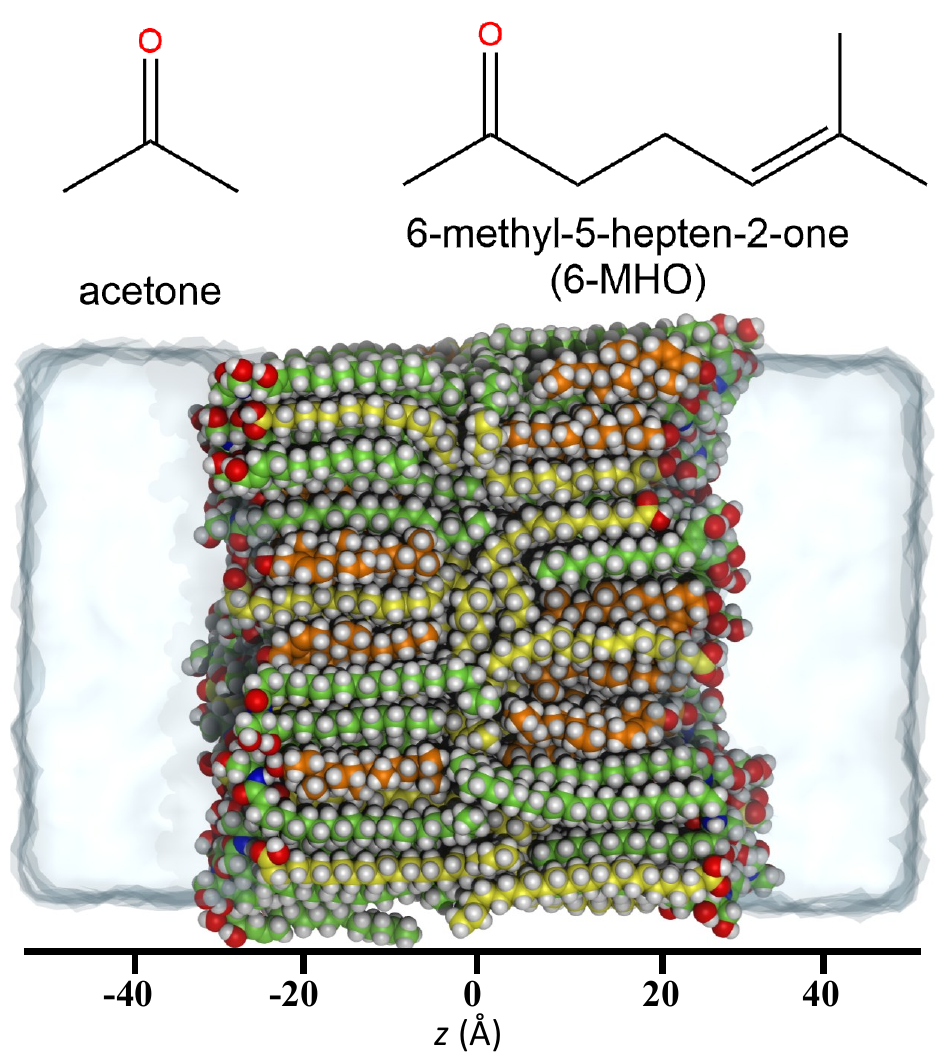}
    \caption{
        Chemical structures of acetone and 6-MHO, and a representative snapshot of the model system. 
        The atomistic lipid membrane is composed of cholesterol ({orange}), lignoceric acid ({yellow}), and ceramide~NS ({green}). 
        In this study, we investigate solute permeation across the membrane along the $z$-axis.
        Figure adapted with permission from ref.~\cite{thomas_insights_2025}.
    }
    \label{fig:model}
\end{figure}

In this study, we address this challenge by investigating the permeation of two representative volatile and semivolatile organic compounds~(VOCs) through human skin. 
The compounds of interest—acetone and 6-MHO (Figure~\ref{fig:model})—are formed during the ozonolysis of skin lipids, which act as the dominant ozone scavengers in densely occupied indoor environments and thus constitute an important source of secondary indoor pollutants.\cite{Wisthaler.Weschler.2010.ReactionsOzoneHuman,Weschler.Weschler.2016.RolesHumanOccupant}
Kinetic models have been developed to describe the mass transport and chemical transformations of these oxidation products across the skin, surrounding air, and clothing.\cite{Lakey.Shiraiwa.2017.ChemicalKineticsMultiphase,Lakey.Shiraiwa.2019.ImpactClothingOzone}
However, the reliability of such models depends critically on accurate physicochemical parameters describing the partitioning of semivolatile compounds—such as 6-MHO—between the human body and indoor air,\cite{VonDomaros.Tobias.2020.MultiscaleModelingHuman} as well as on reliable descriptions of their diffusional transport through the skin barrier.

To address this challenge, we recently began investigating permeation energetics using atomistic molecular dynamics~(MD) simulations of stratum corneum~(SC) lipid membranes ({Fig~\ref{fig:model}})—the skin’s primary barrier to chemical penetration.\cite{thomas_insights_2025}
We found that the convergence of free-energy~(FE) profiles describing the translocation of permeants across these membranes was remarkably slow, requiring microseconds of simulation time even when state-of-the-art enhanced sampling techniques were employed.
We traced the molecular origin of this convergence bottleneck to the highly ordered structure of SC membranes, which contrasts sharply with the fluid-like organization of conventional lipid bilayers such as those composed of 1-palmitoyl-2-oleoyl-\emph{sn}-glycero-3-phosphocholine~(POPC).
In addition, we identified stochastic lipid flip–flop events between the two membrane leaflets that induce long-lived structural asymmetries.

In this work, we extend our previous investigations by providing estimates of diffusivity profiles along the membrane normal. Predicting these diffusivities remains a challenging task—particularly when permeation is slow and enhanced sampling is required to overcome large FE barriers. We employ two complementary approaches to compute position-dependent diffusivities from harmonically restrained MD simulations, based on either the velocity autocorrelation function~(VACF)\cite{Woolf.Roux.1994.ConformationalFlexibilityOPhosphorylcholine} or the position autocorrelation function~(PACF)\cite{Hummer.Hummer.2005.PositiondependentDiffusionCoefficients} of the coordinate describing membrane permeation. 
As shown by our results—and consistent with the findings of Rowley \textit{et~al.} for a dipalmitoylphosphatidylcholine~(DPPC) membrane\cite{gaalswyk_generalized_2016}—estimates obtained from the two approaches differ substantially. Through a rigorous propagator-based analysis, we demonstrate that the resulting diffusivity profiles provide lower and upper bounds to the true profile. 
We further propagate these results to obtain single-membrane permeabilities and we critically assess how methodological uncertainties affect the accuracy of these predictions. Taken together, our work advances the understanding of molecular permeation across skin lipid membranes and establishes a robust framework for improving predictive models of indoor air chemistry under changing environmental conditions.

\section{Methods}

We use atomistic MD simulations, which follow the motion of all atoms in the system in time according to classical mechanics.
This approach allows us to quantify how individual molecules move through a model skin–lipid barrier and how readily they cross it.
The resulting microscopic transport properties serve as input parameters for larger-scale models of skin transport and indoor air chemistry.
The model systems and simulation protocol have been described in detail in our previous study.\cite{thomas_insights_2025} 
Here, we briefly summarize the main aspects relevant to the present work and describe the computation of position-dependent diffusivities and the propagator analysis.

\subsection{Model Systems}

The SC is a complex structure composed of cornified cells embedded in a lipid matrix, an arrangement often described as a brick-and-mortar architecture.\cite{Michaels.Shaw.1975.DrugPermeationHuman, Elias.Elias.1983.EpidermalLipidsBarrier,Schmitt.Neubert.2020.StateArtStratum}
Atomistic studies of skin permeation typically focus on the lipid matrix, which is the only continuous component of the SC and represents a major barrier to permeation.
Extrapolation to the full SC requires combining MD-based transport properties with macroscopic models that account for additional SC components and the detailed permeation pathway.\cite{Wang.Nitsche.2007.MultiphaseMicroscopicDiffusion,Wang.Nagel.2023.TransdermalLateralEffective}

The structure of the SC lipid matrix has been extensively characterized using a range of experimental techniques. There is broad consensus that it consists predominantly of an approximately equimolar mixture of ceramides (CERs), free fatty acids (FFAs), and cholesterol (CHOL), organized into at least two lamellar microphases: the short periodicity phase (SPP) and the long periodicity phase (LPP).\cite{Sparr.Topgaard.2023.StratumCorneumBarrier,Bouwstra.Gooris.2023.SkinBarrierExtraordinary}

The SPP, with a repeat distance of about \SIrange{5}{6}{nm}, is commonly described as a stack of lipid bilayers.
In contrast, the LPP, with a repeat distance of roughly \SIrange{12}{14}{nm}, is often regarded as a more complex multilamellar arrangement stabilized by $\omega$-acyl ceramides that may adopt extended rather than hairpin conformations.
Substantial modeling efforts have been dedicated to understanding both phases, with recent work emphasizing the LPP, whose precise molecular structure remains incompletely resolved.\cite{Sparr.Topgaard.2023.StratumCorneumBarrier,Bouwstra.Gooris.2023.SkinBarrierExtraordinary}

Here, we investigate the permeation of three solutes—acetone, 6-MHO, and water—representing two skin-oil oxidation products and a reference molecule.
These solutes permeate through a model SC lipid bilayer originally proposed by Wang and Klauda\cite{wang_models_2018} and representative of the SPP. The present work focuses on the SPP as a tractable and widely used model system. Establishing robust methodology for diffusivity estimation in this simpler context provides a foundation for future studies that will incorporate the LPP and more realistic SC architectures.

\subsection{Umbrella Sampling}

The translocation of the solutes across the membrane is described by a collective variable~(CV), denoted as $z$, which measures the center-of-mass distance between each solute and the membrane projected onto the membrane normal ({Figure~\ref{fig:model}}).
The CV $z$ can be viewed as a one-dimensional reaction coordinate that tracks how far the molecule has moved from the membrane center toward either side.
Instead of relying on a single long simulation to capture rare permeation events, we divide the coordinate into many overlapping windows along $z$, where each window focuses the sampling on a narrow region of the membrane, in a procedure known as umbrella sampling (US).
The solute is confined within each window by a harmonic potential, which permits fluctuations around a chosen position while still allowing the solute to explore the local environment.
For each solute, we performed 91 US windows, in which the solute was harmonically restrained at integer-\si{\angstrom} positions along the $z$ coordinate.
 
A detailed analysis of the influence of the harmonic force constant was presented by Rowley \textit{et~al.},\cite{gaalswyk_generalized_2016} and our own related findings on this topic are provided in {Section~S1} of the Supporting Information~(SI).
For the results reported in the main text, a force constant of \SI{30}{\kcal\per\mol\per\angstrom\squared} is employed. 
Each simulation is initiated from well-equilibrated configurations obtained in our previous study and run for \SI{15}{\ns} per window.
Although simulation times per window are short, the combined sampling over all windows effectively corresponds to many independent permeation attempts across the entire membrane.
All simulations are carried out at a temperature close to that of the outer skin surface ($T = \SI{305.15}{\kelvin}$) to mimic realistic physiological conditions.
The temperature is maintained by a stochastic velocity-rescaling thermostat\cite{bussi_canonical_2007} with a time constant of \SI{1}{\ps}.
All simulations are performed using NAMD~3.0 (alpha13).\cite{phillips_scalable_2005,phillips_scalable_2020}

\subsection{Diffusivity Profiles}

In addition to the FE barriers, a key quantity for transport is how quickly a molecule can move at different depths within the membrane.
We express this by a position-dependent diffusivity $D(z)$, which may be interpreted as a local mobility: large $D(z)$ indicates that the molecule can move easily at that position, whereas small $D(z)$ reflects strong hindrance by the surrounding lipids.
Two approaches are used to compute the position-dependent diffusivities, $D(z)$, along the membrane normal.

The first approach, proposed by Woolf and Roux, relies on the calculation of the velocity autocorrelation function~(VACF):\cite{Woolf.Roux.1994.ConformationalFlexibilityOPhosphorylcholine}
\begin{equation}
C_v(t) = \left<\dot{z}(t)\dot{z}(0)\right>.
\end{equation}
The VACF quantifies how quickly the solute’s velocity loses memory of its initial direction of motion.
If a molecule continues to move in approximately the same direction for a long time, it is effectively more mobile than if its velocity becomes randomized on a short timescale.
To extract a position-dependent diffusivity $D(z)$, one must perform a Laplace transform of the VACF, $\hat{C}_v(s)$, and extrapolate the resulting Laplace-frequency–dependent diffusivity,
\begin{equation}
D(s) = 
\frac{
    -\hat{C}_v(s)\left<\delta z^2\right>\left<\dot{z}^2\right>
}{
    \hat{C}_v(s)\!\left[s\!\left<\delta z^2\right> + s^{-1}\!\left<\dot{z}^2\right>\right] 
    - \left<\delta z^2\right>\left<\dot{z}^2\right>
},
\label{eq:ds}
\end{equation}
to the limit of zero frequency (infinite lag time),
\begin{equation}
D(z=\left<z\right>) = \lim_{s \to 0} D(s).
\label{eq:ds_extrapol}
\end{equation}
However, this limit is ill-defined because of the $s^{-1}$ dependence and therefore requires an empirical extrapolation procedure.

Hummer later proposed an alternative approach that avoids Laplace-domain extrapolation by using the position autocorrelation function~(PACF):\cite{Hummer.Hummer.2005.PositiondependentDiffusionCoefficients}
\begin{align}
C_{\delta z}(t) &= \frac{\left<\delta z(t)\delta z(0)\right>}{\left<\delta z^2\right>}, \\
D(z=\left<z\right>) &= \frac{\left<\delta z^2\right>}{\int_0^\infty C_{\delta z}(t) \, dt}.
\label{eq:dt}
\end{align}
The PACF characterizes how strongly the current position of the solute along $z$ remains correlated with its initial position.
In regions where the molecule moves freely, this positional correlation decays rapidly, whereas in regions where its motion is constrained, the correlation persists for longer times.
In practice, however, one must either impose a cutoff time for the integral or assume a functional form to fit the correlation decay—both of which can introduce systematic bias.

The velocities required for computing the VACF are obtained from finite differences of the CV trajectory, which is saved at \SI{2}{\femto\second} intervals. 
Position-dependent diffusivities are determined using the {ACFCalculator} code developed by Rowley \textit{et~al.}, which implements a heuristic procedure for the extrapolation in eq.~\ref{eq:ds_extrapol}.\cite{gaalswyk_generalized_2016}

We note that alternative strategies for estimating position-dependent diffusivities are available.
In particular, Bayesian-inference methods, which were originally proposed by Hummer\cite{hummer_position-dependent_2005} 
and later extended to anisotropic media by Ghysels \textit{et al.}\cite{Ghysels.Hummer.2017.PositionDependentDiffusionTensors}, have also been developed.
They have also been coupled to adaptive biasing force calculations by Comer, Chipot, and co-workers.\cite{Comer.Gonzalez-Nilo.2013.CalculatingPositionDependentDiffusivity,Chipot.Comer.2016.SubdiffusionMembranePermeation}
Complementary formulations have been developed by Rosta and collaborators.\cite{Rosta.Hummer.2015.FreeEnergiesDynamic,Sicard.Rosta.2021.PositionDependentDiffusionBiased} 
Position-dependent diffusivities can also be estimated from the autocorrelation function of the constraint force (FACF) in simulations where the particle is algorithmically constrained to selected positions $z$ along the CV. However, this approach can be sensitive to finite-size and momentum-conservation artifacts and therefore requires particular care in its practical implementation.\cite{FujimotoNagaiEtAl2021}
We focus here on the VACF/PACF approaches because they are widely used in membrane permeation studies,\cite{das_water_2009,gupta_molecular_2016,kramer_membrane_2020,piasentin_evaluation_2021,Reuter.Lunter.2025.PresenceDifferentCeramide} and they allow direct comparison with prior results for our benchmark system (water).

Using both VACF- and PACF-based estimators on the same simulation data allows us to assess how sensitive the inferred diffusivities are to the choice of analysis method.
As shown below, the two approaches agree well in the bulk water regions but diverge near the membrane center, where the dynamics is more complex.
We interpret these differences as providing a realistic range (lower and upper bounds) for the true local diffusivity.

\subsection{Propagator Analysis}

Diffusion in inhomogeneous media can be described by the Smoluchowski equation. 
In the one-dimensional case relevant to this study, the position- and time-dependent probability density $p(z, t)$ evolves according to
\begin{equation}
    \partial_t p(z, t) = \partial_z \left\lbrace 
        D(z) e^{-\beta \Delta F(z)} \, \partial_z \!\left[
            e^{\beta \Delta F(z)} p(z, t)
        \right]
    \right\rbrace ,
\end{equation}
where $\beta = 1/(k_\mathrm{B}T)$ is the inverse thermal energy, $\Delta F(z)$ denotes the FE profile, and $D(z)$ represents the position-dependent diffusivity.
This one-dimensional Smoluchowski equation describes diffusion on a free-energy landscape: the molecule tends to move from regions of high free energy to regions of lower free energy, while its motion is locally slowed down or accelerated by the position-dependent diffusivity $D(z)$.
In this formulation, the thermodynamic driving forces are encoded in $\Delta F(z)$ and the kinetic hindrance in $D(z)$, and both contributions enter a single evolution equation for the probability of finding the molecule at a given position and time.

For given $\Delta F(z)$ and $D(z)$ profiles and a delta-function initial condition, $p(z, t_0) = \delta(z - z_0)$, this equation can be solved numerically using a standard forward-in-time, centered-in-space~(FTCS) finite-difference scheme. 
The resulting solution, $p(z, t \,|\, z_0, t_0)$, represents the propagator of the diffusion model—that is, the conditional probability of finding a particle initially located at $z(t_0) = z_0$ at position $z(t)$ at a later time $t$. 
The same quantity can also be obtained directly from molecular dynamics~(MD) simulations without assuming diffusive behavior, thereby providing a rigorous test of the accuracy of the proposed $\Delta F(z)$ and $D(z)$ profiles.
Comparing these model-based propagators to those obtained directly from MD simulations provides a stringent test of the validity of the chosen $\Delta F(z)$ and $D(z)$ profiles.

We compute MD-derived propagators for the critical membrane-core region as follows. 
A total of 1800 snapshots were selected in which the solute was located near the membrane center ($|z| < \SI{0.05}{\angstrom}$). The harmonic restraint was then removed, allowing the solute to freely explore the entire membrane. Unrestrained simulations were run for \SI{15}{\nano\second}, and propagators $p(z, t \,|\, z_0 = 0, t_0 = 0)$ for selected lag times $t$ were estimated from histograms along the $z$ coordinate.

\section{Results and Discussion}

\subsection{Diffusivity Profiles}

\begin{figure*}
    \centering
    \includegraphics[width=\textwidth]{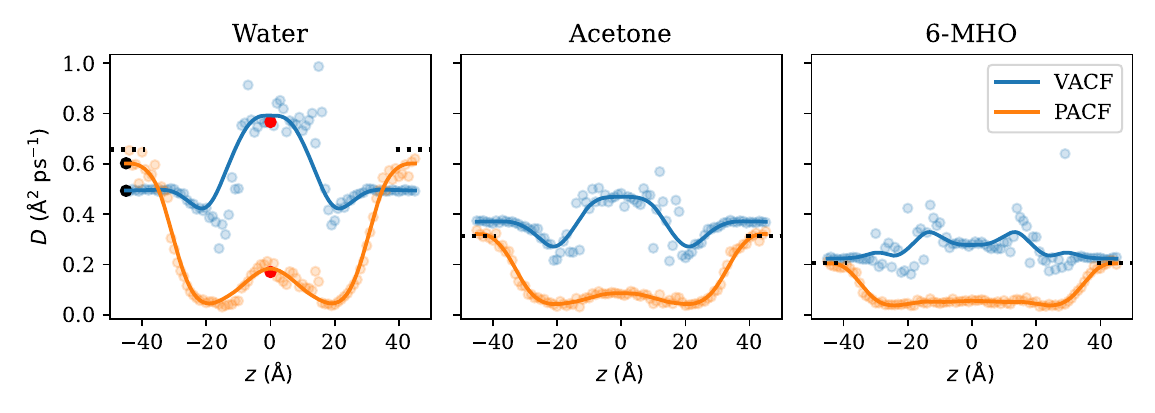}
    \caption{
        Transmembrane diffusivity profiles obtained from the velocity autocorrelation function~(VACF) and position autocorrelation function~(PACF) analyses. 
        Quantitative and qualitative discrepancies are evident near the membrane center. 
        Solid lines are shown to guide the eye, and dashed lines indicate reference values for the aqueous-phase bulk diffusivities. 
        In the water profile, two representative data points—one in the bulk region (black) and one at the membrane center (red)—are highlighted for further analysis.
    }
    \label{fig:diffusivities}
\end{figure*}

Diffusivity profiles, $D(z)$, of the solutes along the transmembrane axis were estimated from the same set of simulations using the PACF and VACF approaches described above. The resulting profiles are shown in Figure~\ref{fig:diffusivities}.

For all three systems, the PACF method yields constant diffusivities in the aqueous regions (\(|z| > \SI{40}{\angstrom}\)) surrounding the membrane, which agree well with the corresponding bulk-phase diffusivities obtained from separate simulations ({Section~S2}). 
The VACF method produces comparable results in this region for acetone and 6-MHO but deviates by approximately \SI{24}{\percent} for water.

In the headgroup region, both methods indicate a local diffusivity bottleneck for all solutes. 
The most pronounced discrepancy between the two approaches, however, appears at the membrane center, where the estimates differ both quantitatively and qualitatively: the VACF method predicts higher-than-bulk diffusivities, whereas the PACF method yields lower values. 
Similar observations were reported by Rowley \textit{et~al.}\ for a DPPC membrane.\cite{gaalswyk_generalized_2016} 
In the following, we discuss the challenges associated with determining diffusivities using both approaches and the likely origins of these discrepancies. 
Our analysis focuses primarily on two representative regions of the membrane: the aqueous region, where the two methods show good agreement, and the membrane center, where the discrepancies are most pronounced.

\subsection{Challenge 1: Extrapolation of the VACF Results}

\begin{figure}
    \centering
    \includegraphics[width=\columnwidth]{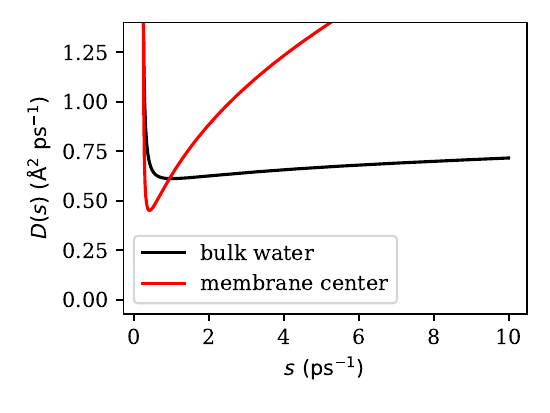}
    \caption{
        Illustration of determining diffusivities from velocity autocorrelation functions. 
        The quantity $D(s)$, computed from eq.~\ref{eq:ds}, must be extrapolated to zero by fitting a straight line over a range of $s$ values where the function remains well-behaved. 
        This procedure performs reliably in the bulk phase but is highly sensitive to the placement of the tangent line for windows at the membrane center. 
        The values obtained using the automated extrapolation procedure of Rowley \textit{et~al.}\cite{gaalswyk_generalized_2016} are indicated by the black and red points in Figure~\ref{fig:diffusivities}.
    }
    \label{fig:vacf_extrapolation}
\end{figure}

Obtaining position-dependent diffusivities from VACFs requires computing a Laplace-frequency–dependent diffusivity, $D(s)$, for each US window (eq.~\ref{eq:ds}), followed by extrapolation to zero frequency (eq.~\ref{eq:ds_extrapol}). 
However, this limit is ill-defined because of the $s^{-1}$ dependence of $D(s)$ and therefore necessitates an empirical extrapolation procedure. Our VACF-based diffusivity profiles were obtained using the automated linear extrapolation method of Rowley \textit{et al.}\cite{gaalswyk_generalized_2016} Although this approach is relatively robust in the aqueous bulk region, it may not be optimal at the membrane center (Figure~\ref{fig:vacf_extrapolation}). Because $D(s)$ exhibits pronounced curvature in the membrane core, the linear extrapolation is highly sensitive to the placement of the tangent line, yielding diffusivity estimates between roughly \SI{0.1}{\angstrom^2\per\pico\second} and \SI{0.75}{\angstrom^2\per\pico\second}.

\subsection{Challenge 2: Determination of PACF Correlation Times}

\begin{figure}
    \centering
    \includegraphics[width=\columnwidth]{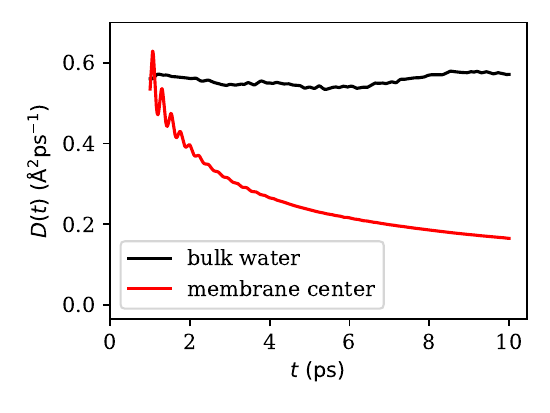}
    \caption{
        Illustration of determining diffusivities from position autocorrelation functions~(PACFs). 
        Shown is the estimated diffusivity, $D(t)$, as a function of the PACF integration cutoff time in two selected regions of the membrane. 
        In the absence of noise, $D(t)$ should reach a plateau. 
        A clear plateau is visible in the bulk water region but not at the membrane center. 
        The values corresponding to a fixed cutoff of $t = \SI{10}{\pico\second}$ are indicated by the black and red points in Figure~\ref{fig:diffusivities}.
    }
    \label{fig:pacf_extrapolation}
\end{figure}

The challenges associated with extrapolating $D(s)$ to zero frequency were recognized by Hummer, who proposed an alternative approach based on determining the PACF correlation time (eq.~\ref{eq:dt}).\cite{hummer_position-dependent_2005} 
For purely diffusive dynamics, this method is formally equivalent to the VACF approach, but it is considerably simpler to apply in practice.

In Figure~\ref{fig:pacf_extrapolation}, we show selected diffusivities, $D(t)$, as a function of the integration cutoff time used in eq.~\ref{eq:dt}. 
The PACF method performs well when the solute resides in the aqueous phase, where $D(t)$ reaches a clear plateau between approximately \SI{8}{\ps} and \SI{10}{\ps}. 
Although noise leads to some fluctuations, it does not significantly alter the predicted diffusivity. 
In contrast, at the membrane center, the decay of the PACF is substantially slower, and no clear plateau in $D(t)$ is observed even after \SI{10}{\ps}. 
These observations are consistent with those reported by Rowley \textit{et~al.}\ for a DPPC membrane.\cite{gaalswyk_generalized_2016}

As a consequence, solute- and region-specific integration cutoffs would be ideal but require tedious manual adjustment. Instead, the diffusivity profiles in Figure~\ref{fig:diffusivities} were obtained using practical compromise cutoffs: \SI{8}{\ps} for water and \SI{10}{\ps} for acetone and 6-MHO. Although these fixed cutoffs may slightly overestimate PACF-derived diffusivities in certain regions (e.g., the membrane center), our main conclusions are unaffected: PACF-derived profiles still represent lower bounds to the true $D(z)$.

\subsection{Propagator Analysis}

\begin{figure}
    \centering
    \includegraphics[width=\columnwidth]{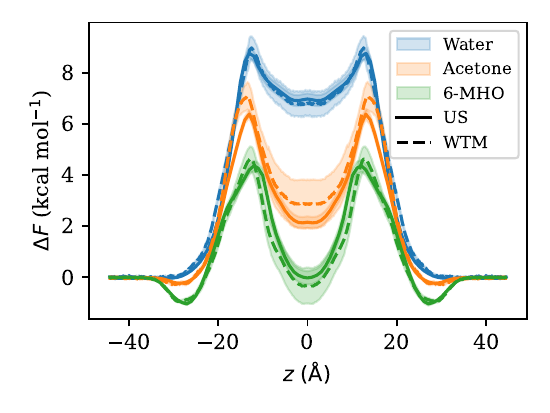}
    \caption{
        Free-energy profiles for the translocation of all solutes across the model stratum corneum membrane, obtained from umbrella sampling~(US) and well-tempered metadynamics~(WTM).\cite{thomas_insights_2025} 
        The two methods produce consistent overall features, with only minor quantitative differences. 
        Figure adapted with permission from ref.~\cite{thomas_insights_2025}.
    }
    \label{fig:free_energies}
\end{figure}

\begin{figure*}
    \centering
    \includegraphics[width=\textwidth]{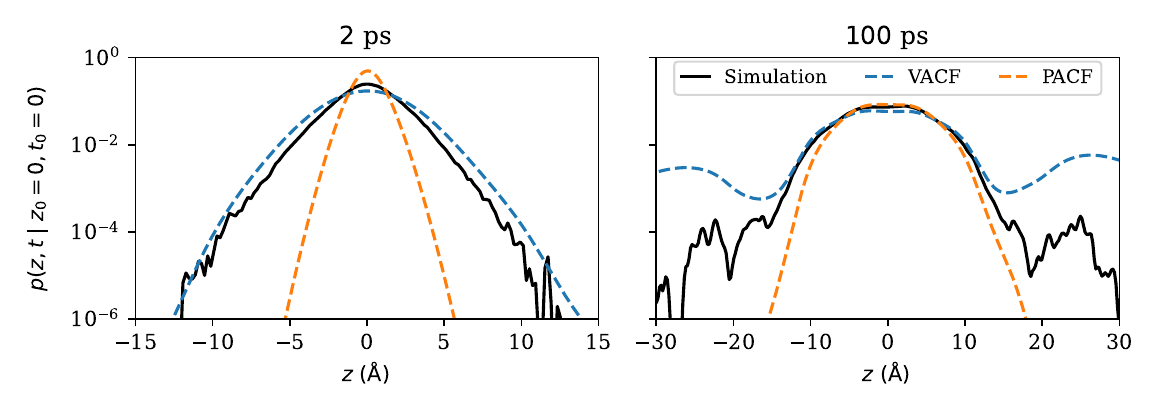}
    \caption{
        Propagators, $p(z, t \,|\, z_0, t_0)$, representing the probability of finding a particle initially located at $z_0 = 0$ at position $z$ after a lag time $t$. 
        Shown are results obtained by numerically solving the Smoluchowski diffusion equation using diffusivity profiles derived from the velocity autocorrelation function~(VACF) and position autocorrelation function~(PACF) analyses, compared with direct simulation data. 
        At short lag times ($t = \SI{2}{\pico\second}$), the VACF-based model shows quantitative agreement with the simulation data, although the true diffusive regime has likely not yet been reached on this timescale. 
        At longer lag times ($t = \SI{100}{\pico\second}$), the VACF- and PACF-based models provide upper and lower bounds, respectively, to the true propagators and, by extension, to the diffusivity profile.
    }
    \label{fig:propagators}
\end{figure*}

Having obtained two qualitatively distinct diffusivity profiles from the same set of simulations, it is natural to ask which of the two, if either, is correct. 
To address this question, we computed the propagators of both diffusion models in the critical membrane-center region, $p(z, t \,|\, z_0 = 0, t_0 = 0)$—that is, the probability of finding a particle initially located at $z_0 = 0$ at position $z$ after a lag time $t$. 
The corresponding FE profiles were taken from our previous study,\cite{thomas_insights_2025} as shown in Figure~\ref{fig:free_energies}. 
For the present analysis, we used the umbrella-sampling (US)–based free-energy profiles. 
A comparison between the model-derived propagators and those obtained directly from molecular dynamics simulations is presented in Figure~\ref{fig:propagators} for two representative lag times.

At short lag times ($t = \SI{2}{\pico\second}$), the VACF-based model shows quantitative agreement with the simulation data over six orders of magnitude, whereas the PACF approach fails to do so. 
This finding is consistent with the results reported by Rowley \textit{et~al.},\cite{gaalswyk_generalized_2016} who performed a simplified analysis by approximating the propagator as a normal distribution—an assumption that holds only for lag times long enough to reproduce diffusive dynamics yet short enough that the effects of inhomogeneity remain negligible.

However, it is not obvious that the lag time of \SI{2}{\pico\second} chosen here (or the \SI{5}{\pico\second} used by Rowley \textit{et~al.}) is sufficiently long for the diffusive approximation to hold. 
Indeed, the failure of the PACF to decay within such short timescales may also be interpreted as a breakdown of this assumption. 
Subdiffusive behavior has, in fact, been reported for lag times up to at least \SI{64}{\pico\second} for methanol permeation across a POPC membrane.\cite{Chipot.Comer.2016.SubdiffusionMembranePermeation}

We therefore extended our propagator analysis to selected lag times of up to $\SI{500}{\pico\second}$. 
For a range of lag times between {\SI{50}{\pico\second}} and {\SI{500}{\pico\second}}, which we consider to lie within the truly diffusive regime, the model propagators exhibit the behavior shown for the representative case of $t = \SI{100}{\pico\second}$ in Figure~\ref{fig:propagators}. 
The complete set of results is provided in {Section~S3}.

At these timescales, neither the VACF- nor the PACF-based models reproduce the simulation data quantitatively. 
However, the VACF results consistently overestimate the diffusive spread (i.e., the diffusivities are too large), whereas the PACF results underestimate it (i.e., the diffusivities are too small). 
We therefore interpret the two profiles as providing upper and lower bounds, respectively, to the true diffusivity profile, which must lie between them.
In the following, these bounds are used to assess the sensitivity of derived quantities, specifically the transmembrane permeabilities.

\subsection{Impact on Permeabilities}

\begin{figure}
    \centering
    \includegraphics[width=\columnwidth]{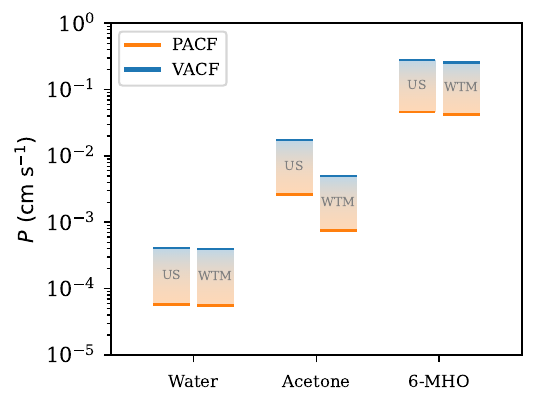}
    \caption{
        Membrane permeabilities estimated using diffusivity profiles derived from the velocity autocorrelation function~(VACF, left) and position autocorrelation function~(PACF, right) analyses, each combined with free-energy profiles obtained from umbrella sampling~(US) and well-tempered metadynamics~(WTM). 
        Differences between the free-energy methods are noticeable only for acetone, whereas discrepancies between the two diffusivity analyses consistently span approximately one order of magnitude across all solutes.
    }
    \label{fig:permeabilities}
\end{figure}

Within the inhomogeneous solubility--diffusion (ISD) model,\cite{Diamond.Katz.1974.InterpretationNonelectrolytePartition,Marrink.Berendsen.1994.SimulationWaterTransport} 
the local resistance of a membrane to permeation, $R(z)$, is expressed in terms of the diffusivity and FE profiles as
\begin{equation}
    R(z) = \frac{e^{\beta \Delta F(z)}}{D(z)} \, .
    \label{eq:resistance}
\end{equation}
The overall permeability of the membrane, $P$, is then obtained by integrating the resistance over the membrane width, $h$, according to
\begin{equation}
    \frac{1}{P} = \int_{-h/2}^{+h/2} R(z) \, dz .
    \label{eqn::perm}
\end{equation}

Using the FE profiles for the translocation of all solutes across the membrane (Figure~\ref{fig:free_energies}), we computed the membrane permeabilities according to eqs~\ref{eq:resistance} and~\ref{eqn::perm}. 
The resulting permeabilities are shown in Figure~\ref{fig:permeabilities}, and the corresponding numerical values are provided in the Supporting Information ({Section~S4}). 
Because of the exponential dependence on the free energy in eq~\ref{eq:resistance}, even small variations in $\Delta F(z)$ are amplified, leading to pronounced differences in the resulting permeabilities, which follow the order 6-MHO~$>$~acetone~$\gg$~water.

We have also shown that the FE profiles obtained from US and WTM differ slightly (Figure~\ref{fig:free_energies}), even after extensive sampling, although these differences remain within the limits of statistical uncertainty.\cite{thomas_insights_2025} 
With the diffusivity profiles now available, we can propagate the different FE profiles to evaluate their impact on the resulting transport coefficients. 
We find good agreement between the results obtained from both methods for water and 6-MHO; for acetone, however, noticeable deviations are observed. 
Nevertheless, the permeability coefficients for all solutes differ by less than an order of magnitude, indicating that statistical variations between FE methods likely have little to no impact on kinetic model predictions.

More striking are the differences in permeabilities when comparing VACF- and PACF-derived diffusivities. 
Because the VACF-derived diffusivities are consistently higher than those obtained from the PACF method, the same trend is reflected in the corresponding permeabilities. 
In absolute terms, deviations reach approximately one order of magnitude. 
Whether such differences are significant for kinetic models is highly problem specific. 
In many cases, these discrepancies are likely acceptable, although the final assessment should be guided by a dedicated sensitivity analysis.\cite{Lakey.Shiraiwa.2019.ImpactClothingOzone}

It is important to note here that diffusivity, free energy, and permeation reflect different aspects of transport. 
A high free-energy barrier (as in the case of water at the bilayer center) reduces the probability of occupancy in that region but does not necessarily imply fast diffusion once a molecule is there.
Permeability depends on both thermodynamics and kinetics through eq.~\ref{eq:resistance}, and intuition based solely on an unfavourable free-energy environment risks conflating the role of $\Delta F(z)$ with that of $D(z)$. 
Diffusivity profiles $D(z)$ probe the mobility of a solute conditioned on being at position $z$, not the likelihood of being there in the first place. 
Within this interpretation, a lower diffusivity $D(z)$ at the membrane center does not mean that water prefers this region, only that its motion is more hindered when present.
A surprising consequence of this interplay between thermodynamic and kinetic effects is the relatively high permeability of 6-MHO, despite the fact that its diffusivity profile is generally the lowest of the three solutes.

Finally, we examined the influence of the integration limits used in eq.~\ref{eqn::perm}. 
Typically, $h$ is defined as the full thickness of the membrane. 
However, for heterogeneous, multicomponent lipid membranes such as the SC system investigated here, determining precise interfacial boundaries ($\pm h/2$) is nontrivial. 
We therefore considered three plausible definitions for $h/2$, based on the membrane density profile ({Section~S4}): 
(\textit{i}) the position where the water density begins to decrease from its bulk value ($h_1/2 = \qty{32.0}{\angstrom}$); 
(\textit{ii}) the position where the water density is reduced to half of its bulk value ($h_2/2 = \qty{25.5}{\angstrom}$); and 
(\textit{iii}) the position where the water density approaches zero ($h_3/2 = \qty{17.0}{\angstrom}$).

\begin{figure*}
    \centering
    \includegraphics[width=\textwidth]{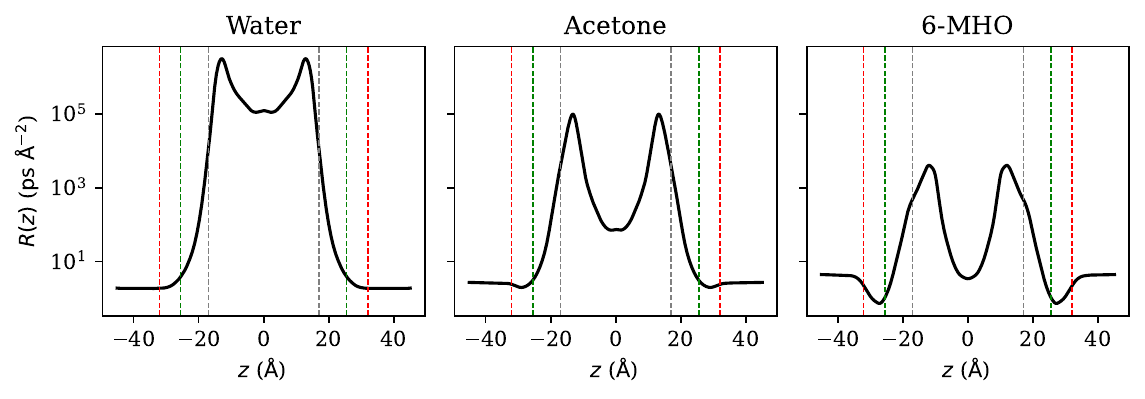}
    \caption{
        Local resistance, $R(z)$, estimated using free-energy profiles obtained from umbrella sampling and diffusivity profiles derived from the velocity autocorrelation function analysis. 
        The corresponding results for other combinations of free-energy and diffusivity methods are provided in {Section~S4}. 
        The red, green, and gray vertical lines indicate different plausible integration limits. 
        The areas under the curves in these regions are negligible compared with the contributions from the membrane center, indicating that different integration limits have little impact on the overall permeability.
    }
    \label{fig:resistance_us_vacf}
\end{figure*}

To investigate the impact of these definitions, we show the local resistance $R(z)$ in Figure~\ref{fig:resistance_us_vacf}. 
The area under each curve corresponds to the inverse permeability, $P^{-1}$. 
Within the membrane interior, this area is several orders of magnitude larger than the small additional contributions from the interfacial regions. 
We therefore conclude that different reasonable integration limits have little to no effect on the computed permeabilities.

We conclude this section by noting that a direct molecular interpretation of the resistance profiles in Fig.~\ref{fig:resistance_us_vacf} in terms of the membrane structure in Fig.~\ref{fig:model} is non-trivial. 
The local resistance $R(z)$ reflects the combined effect of free-energy barriers and diffusivity (eq.~\ref{eq:resistance}), meaning that structural features do not map one-to-one onto the profile but instead contribute through a convolution of thermodynamic and kinetic influences. 
Nevertheless, we can conjecture about some qualitative structure–transport relations.
The elevated resistance near the headgroup region is consistent with the dense packing and strong hydrogen-bonding environment at lipid–water interfaces, which restricts mobility particularly for polar permeants. 
In the membrane center, the loosened lipid organization and resulting elevated free volume give rise to conditions that may accelerate local diffusion and contribute to the central resistance dip.

\begin{table*}

\centering
\caption{
Water permeability coefficients of stratum corneum (SC) lipid membranes, obtained from molecular dynamics simulations and experimental compilations. Simulation results are reported together with the corresponding force fields, temperatures, and analysis methods used to compute permeability, which differ across studies and contribute to variability in the reported values. Units are $\si{\centi\meter\per\second}$.
}
\label{tab:Permeabilities_exp}
\begin{threeparttable}
\begin{tabular*}{\textwidth}{@{\extracolsep{\fill}}llll}
\toprule
Reference & Permeability & Approach & Details \\
\midrule

\multicolumn{4}{@{}l@{}}{Gupta \textit{et al.} \cite{gupta_molecular_2016}} \\
& $7.08 \pm 1.18 \times 10^{-5}$ & FACF & 310 K, GROMOS/SPC \\

\addlinespace
\multicolumn{4}{@{}l@{}}{Das \textit{et al.} \cite{das_water_2009}} \\
& $8.2 \times 10^{-8}$ & FACF & 350 K, OPLS/SPC \\

\addlinespace
\multicolumn{4}{@{}l@{}}{Reuter \textit{et al.} \cite{reuter_presence_2025}} \\
& $1.88 \times 10^{-5}$ & FACF & 303.15 K, CHARMM36/TIP3P \\

\addlinespace
\multicolumn{4}{@{}l@{}}{Piasentin \textit{et al.} \cite{piasentin_evaluation_2021}} \\
& $2.79 \pm 0.52 \times 10^{-5}$ & FACF & 303.15 K, CHARMM36/TIP3P \\
& $2.10 \pm 0.39 \times 10^{-5}$ & US + FACF & 303.15 K, CHARMM36/TIP3P \\
& $1.89 \pm 0.34 \times 10^{-5}$ & US + VACF & 303.15 K, CHARMM36/TIP3P \\

\addlinespace
\multicolumn{4}{@{}l@{}}{This work} \\
& $5.65 \times 10^{-5}$ & US + PACF & 305.15 K, CHARMM36/TIP3P \\
& $5.48 \times 10^{-5}$ & WTM + PACF & 305.15 K, CHARMM36/TIP3P \\
& $4.15 \times 10^{-4}$ & US + VACF & 305.15 K, CHARMM36/TIP3P \\
& $4.00 \times 10^{-4}$ & WTM + VACF & 305.15 K, CHARMM36/TIP3P \\

\addlinespace
\multicolumn{4}{@{}l@{}}{Katritzky \textit{et al.} \cite{katritzky_skin_2006}\tnote{a}} \\
& $2.95 \times 10^{-7}$ & Experimental & \\

\addlinespace
\multicolumn{4}{@{}l@{}}{Chen \textit{et al.} \cite{chen_modeling_2010}\tnote{a}} \\
& $4.79 \times 10^{-7}$ & Experimental & \\

\bottomrule
\end{tabular*}

\begin{tablenotes}
\item[a] The reported permeability values are compiled from multiple secondary sources and do not represent primary experimental measurements.
\end{tablenotes}
\end{threeparttable}
\end{table*}

\section{Conclusions}

We used MD simulations to quantify position-dependent diffusivities, $D(z)$, for water, acetone, and 6-MHO across a representative SC lipid membrane and combined these results with previously reported FE profiles to compute transmembrane permeabilities, $P$.
By applying both PACF- and VACF-based analyses to the same set of US trajectories and validating them through a propagator comparison, we demonstrated that the two $D(z)$ estimates systematically bracket the true diffusivity profile in the diffusive regime, with PACF providing a lower bound and VACF an upper bound.
Permeabilities derived from these bounds consequently span approximately one order of magnitude, whereas differences arising from the choice of FE method (US vs.\ WTM) become negligible once propagated to transport parameters.
Consistent with the exponential dependence of the ISD framework on free energy, the ordering of permeabilities follows the underlying FE landscape rather than the local diffusivities—6-MHO $>$ acetone $\gg$ water—highlighting that permeation across the membrane is governed primarily by energetic barriers rather than by differences in molecular mobility.

To place these results in context, we compare the water permeability coefficients obtained in this work with values reported in previous studies. The permeabilities derived from PACF-based diffusivities are of the same order of magnitude as those reported by Gupta et al., Piasentin et al., and Reuter et al. \cite{gupta_molecular_2016,piasentin_evaluation_2021,Reuter.Lunter.2025.PresenceDifferentCeramide}, namely on the order of $10^{-5}$ \si{\centi\meter\per\second} (see Table \ref{tab:Permeabilities_exp}). Residual discrepancies are most likely attributable to differences in simulation temperature and force-field parametrization. Additional methodological factors—such as system size, thermostat and barostat settings, integration parameters, and overall simulation protocols—may also contribute to variations in the reported permeability coefficients. Importantly, this level of agreement should not be interpreted as a validation of the PACF-based diffusivity profiles. The limitations of the approach discussed here are methodological in nature and are therefore expected to be largely independent of specific computational details.

Direct comparison of permeabilities computed from our MD simulations to experimental values (in vitro\cite{anderson_heterogeneity_1988,anderson_solute_1989} or in vivo\cite{kalia_homogeneous_1996}) is not straightforward because our model represents a simplified SC membrane containing only the short periodicity phase (SPP) of the lipid matrix. We deliberately omitted the more structurally complex long periodicity phase (LPP) to first address methodological challenges that arise even in comparatively simple SC models. Improving the realism of atomistic representations is important, but equally essential is ensuring that these methodological limitations are understood and managed if future permeability estimates are to be more robust. We therefore emphasize the need to quantify uncertainty—whether through error bars, bounding estimates, or credibility intervals—and to evaluate how these propagate into higher-scale models.

Permeability coefficients are often estimated empirically using quantitative structure–activity relationship (QSAR) and continuum diffusion models,\cite{wilschut_estimating_1995,lian_evaluation_2008,Mitragotri.Roberts.2011,Frasch.Barbero.2013,Wang.Naegel.2023} which correlate measured permeabilities with molecular descriptors such as molar mass, partition coefficients, and bulk diffusion constants. Although such models are powerful for interpolation within their training domains, they do not explicitly capture the molecular-scale interactions and dynamic heterogeneity that govern transport across the SC lipid matrix—especially when applied to molecules that fall outside their original parameterization range. 
In contrast, MD simulations capture these effects directly by resolving how solutes interact with and move through the heterogeneous lipid environment. For example, despite its larger molar mass relative to water and acetone, 6-MHO exhibits the highest permeability because of its comparatively low free-energy barriers to permeation (Figure~\ref{fig:free_energies}), underscoring the dominant role of thermodynamic driving forces—rather than molecular size—in determining SC transport.

With respect to indoor air chemistry, several kinetic models—such as KM-BL, KM-FILM, and KM-SUB-Skin-Clothing\cite{Lakey.Shiraiwa.2019.ImpactClothingOzone}—have been developed to describe the dynamics of skin-oil oxidation products across different interfaces. However, these models critically depend on physicochemical parameters that are difficult or impossible to measure experimentally, underscoring the value of molecular-scale simulations like those presented here. For example, Lakey \textit{et al.}\ demonstrated that the performance of their KM-FILM model is highly sensitive to the assumed diffusion constants of surface films,\cite{lakey_kinetic_2021} illustrating how improved microscopic input can directly enhance macroscopic predictive accuracy. 

Taken together, our results provide mechanistic and quantitative constraints—via upper and lower bounds on $D(z)$ and permeabilities—that can be directly used to refine existing indoor-air kinetic models.
Beyond these applications, the present framework establishes a foundation for systematically linking atomistic transport processes to macroscopic exposure and air-quality models.
Future directions include (i) extending the approach to multilamellar SC architectures and anisotropic membrane transport; (ii) assessing the influence of force-field and water-model choice; (iii) developing more robust and transferable diffusivity estimators; and (iv) performing targeted experimental validation for VOCs of indoor relevance.
Collectively, these efforts will help bridge the scale gap between molecular transport and human exposure in complex indoor environments.

\begin{acknowledgement}
The authors thank Chinmay Das and Peter~D.~Olmsted for helpful discussions on setting up SC models. DJT acknowledges support from the Alfred P. Sloan Foundation Chemistry of the Indoor Environment (CIE) Program (G-2020-13912).
\end{acknowledgement}

\begin{suppinfo}
The supporting information includes an assessment of the influence of the harmonic force constant on diffusivity profiles; details on the calculation of bulk diffusivities from mean-squared displacements; the complete set of propagator analyses across different lag times; density profiles used to define membrane boundaries; additional local resistance profiles obtained from all combinations of free-energy and diffusivity methods; and a table summarizing all computed permeability coefficients. All structure, topology, force-field, and simulation input files necessary to reproduce the molecular dynamics simulations presented in this work are available in a public GitHub repository at \url{https://github.com/mvondomaros-lab/sc-permeation-est-air}.
\end{suppinfo}

\bibliography{main}

@article{Abdul-Nabi.Zahran.2025.ClimateChangeIts,

  author       = {Abdul-Nabi, Sarah S. and Karaki, Victoria Al and Khalil, Aline and Zahran, Tharwat El},
  title        = {Climate change and its environmental and health effects from 2015 to 2022: A scoping review},
  journal      = {Heliyon},
  year         = {2025},
  month        = feb,
  volume       = {11},
  number       = {3},
  doi          = {10.1016/j.heliyon.2025.e15616},
  pmid         = {39975822}
}

@article{Chipot.Comer.2016.SubdiffusionMembranePermeation,

  author       = {Chipot, Christophe and Comer, Jeffrey},
  title        = {Subdiffusion in membrane permeation of small molecules},
  journal      = {Sci. Rep.},
  year         = {2016},
  volume       = {6},
  number       = {1},
  pages        = {35913},
  doi          = {10.1038/srep35913}
}

@article{Connolly.Connolly.2018.ClimateChangeAllocation,

  author       = {Connolly, Marie},
  title        = {Climate change and the allocation of time},
  journal      = {IZA World Labor},
  year         = {2018},
  month        = jan,
  doi          = {10.15185/izawol.423}
}

@article{Diamond.Katz.1974.InterpretationNonelectrolytePartition,

  author       = {Diamond, Jared M. and Katz, Yehuda},
  title        = {Interpretation of nonelectrolyte partition coefficients between dimyristoyl lecithin and water},
  journal      = {J. Membr. Biol.},
  year         = {1974},
  month        = dec,
  volume       = {17},
  number       = {1},
  pages        = {121--154},
  doi          = {10.1007/BF01870235}
}

@article{Hummer.Hummer.2005.PositiondependentDiffusionCoefficients,

  author       = {Hummer, Gerhard},
  title        = {Position-dependent diffusion coefficients and free energies from Bayesian analysis of equilibrium and replica molecular dynamics simulations},
  journal      = {New J. Phys.},
  year         = {2005},
  month        = jan,
  volume       = {7},
  pages        = {34},
  doi          = {10.1088/1367-2630/7/1/034}
}

@techreport{InstituteOfMedicine.InstituteOfMedicine.2011.ClimateChangeIndoor,

  author       = {{Institute of Medicine}},
  title        = {Climate change, the indoor environment, and health},
  institution  = {National Academies Press},
  address      = {Washington, D.C.},
  year         = {2011},
  month        = aug
}

@incollection{IntergovernmentalPanelOnClimateChangeIPCC.IntergovernmentalPanelOnClimateChangeIPCC.2022.Impacts15degCGlobal,

  editor       = {{Intergovernmental Panel on Climate Change (IPCC)}},
  title        = {Impacts of 1.5$^\circ$C global warming on natural and human systems},
  booktitle    = {Global warming of 1.5$^\circ$C: IPCC special report on impacts of global warming of 1.5$^\circ$C above pre-industrial levels in context of strengthening response to climate change, sustainable development, and efforts to eradicate poverty},
  publisher    = {Cambridge University Press},
  address      = {Cambridge},
  year         = {2022},
  pages        = {175--312},
  doi          = {10.1017/9781009157940.005},
  isbn         = {978-1-009-15795-7}
}

@article{Lakey.Shiraiwa.2017.ChemicalKineticsMultiphase,

  author       = {Lakey, P. S. J. and Wisthaler, A. and Berkemeier, T. and Mikoviny, T. and P{\"o}schl, U. and Shiraiwa, M.},
  title        = {Chemical kinetics of multiphase reactions between ozone and human skin lipids: Implications for indoor air quality and health effects},
  journal      = {Indoor Air},
  year         = {2017},
  volume       = {27},
  number       = {4},
  pages        = {816--828},
  doi          = {10.1111/ina.12360}
}

@article{Lakey.Shiraiwa.2019.ImpactClothingOzone,

  author       = {Lakey, Pascale S. J. and Morrison, Glenn C. and Won, Youngbo and Parry, Krista M. and von Domaros, Michael and Tobias, Douglas J. and Rim, Donghyun and Shiraiwa, Manabu},
  title        = {The impact of clothing on ozone and squalene ozonolysis products in indoor environments},
  journal      = {Commun. Chem.},
  year         = {2019},
  month        = may,
  volume       = {2},
  number       = {1},
  pages        = {1--8},
  doi          = {10.1038/s42004-019-0157-4}
}

@article{Mansouri.Blondeau.2022.ImpactClimateChange,

  author       = {Mansouri, Aya and Wei, Wenjuan and Alessandrini, Jean-Marie and Mandin, Corinne and Blondeau, Patrice},
  title        = {Impact of climate change on indoor air quality: A review},
  journal      = {Int. J. Environ. Res. Public Health},
  year         = {2022},
  month        = jan,
  volume       = {19},
  number       = {23},
  pages        = {15616},
  doi          = {10.3390/ijerph192315616}
}

@article{Marrink.Berendsen.1994.SimulationWaterTransport,

  author       = {Marrink, Siewert-Jan and Berendsen, Herman J. C.},
  title        = {Simulation of water transport through a lipid membrane},
  journal      = {J. Phys. Chem.},
  year         = {1994},
  month        = apr,
  volume       = {98},
  number       = {15},
  pages        = {4155--4168},
  doi          = {10.1021/j100066a040}
}

@article{Nazaroff.Nazaroff.2013.ExploringConsequencesClimate,

  author       = {Nazaroff, William W.},
  title        = {Exploring the consequences of climate change for indoor air quality},
  journal      = {Environ. Res. Lett.},
  year         = {2013},
  month        = feb,
  volume       = {8},
  number       = {1},
  pages        = {015022},
  doi          = {10.1088/1748-9326/8/1/015022}
}

@article{Rocque.Witteman.2021.HealthEffectsClimate,

  author       = {Rocque, Rhea J. and Beaudoin, Caroline and Ndjaboue, Ruth and Cameron, Laura and Poirier-Bergeron, Louann and Poulin-Rheault, Rose-Alice and Fallon, Catherine and Tricco, Andrea C. and Witteman, Holly O.},
  title        = {Health effects of climate change: An overview of systematic reviews},
  journal      = {BMJ Open},
  year         = {2021},
  month        = jun,
  volume       = {11},
  number       = {6},
  pages        = {e046333},
  doi          = {10.1136/bmjopen-2020-046333},
  chapter      = {Public health}
}

@article{Spengler.Spengler.2012.ClimateChangeIndoor,

  author       = {Spengler, John D.},
  title        = {Climate change, indoor environments, and health},
  journal      = {Indoor Air},
  year         = {2012},
  volume       = {22},
  number       = {2},
  pages        = {89--95},
  doi          = {10.1111/j.1600-0668.2012.00770.x}
}

@article{VonDomaros.Tobias.2020.MultiscaleModelingHuman,

  author       = {von Domaros, Michael and Lakey, Pascale S. J. and Shiraiwa, Manabu and Tobias, Douglas J.},
  title        = {Multiscale modeling of human skin oil-induced indoor air chemistry: Combining kinetic models and molecular dynamics},
  journal      = {J. Phys. Chem. B},
  year         = {2020},
  month        = may,
  volume       = {124},
  number       = {18},
  pages        = {3836--3843},
  doi          = {10.1021/acs.jpcb.0c01184}
}

@article{VonDomaros.Tobias.2025.MolecularDynamicsSimulations,

  author       = {von Domaros, Michael and Tobias, Douglas J.},
  title        = {Molecular dynamics simulations of the interactions of organic compounds at indoor-relevant surfaces},
  journal      = {Annu. Rev. Phys. Chem.},
  year         = {2025},
  volume       = {76},
  pages        = {231--250},
  doi          = {10.1146/annurev-physchem-060224-030816}
}

@article{Weschler.Weschler.2016.RolesHumanOccupant,

  author       = {Weschler, C. J.},
  title        = {Roles of the human occupant in indoor chemistry},
  journal      = {Indoor Air},
  year         = {2016},
  volume       = {26},
  number       = {1},
  pages        = {6--24},
  doi          = {10.1111/ina.12235}
}

@article{White-Newsome.ONeill.2012.ClimateChangeHealth,

  author       = {White-Newsome, Jalonne L. and S{\'a}nchez, Brisa N. and Jolliet, Olivier and Zhang, Zhenzhen and Parker, Edith A. and Dvonch, J. Timothy and O'Neill, Marie S.},
  title        = {Climate change and health: Indoor heat exposure in vulnerable populations},
  journal      = {Environ. Res.},
  year         = {2012},
  month        = jan,
  volume       = {112},
  pages        = {20--27},
  doi          = {10.1016/j.envres.2011.10.008}
}

@article{Wisthaler.Weschler.2010.ReactionsOzoneHuman,

  author       = {Wisthaler, Armin and Weschler, Charles J.},
  title        = {Reactions of ozone with human skin lipids: Sources of carbonyls, dicarbonyls, and hydroxycarbonyls in indoor air},
  journal      = {Proc. Natl. Acad. Sci. U.S.A.},
  year         = {2010},
  month        = apr,
  volume       = {107},
  number       = {15},
  pages        = {6568--6575},
  doi          = {10.1073/pnas.0904498106}
}

@article{Woolf.Roux.1994.ConformationalFlexibilityOPhosphorylcholine,

  author       = {Woolf, Thomas B. and Roux, Benoit},
  title        = {Conformational flexibility of \emph{O}-phosphorylcholine and \emph{o}-phosphorylethanolamine: A molecular dynamics study of solvation effects},
  journal      = {J. Am. Chem. Soc.},
  year         = {1994},
  month        = jun,
  volume       = {116},
  number       = {13},
  pages        = {5916--5926},
  doi          = {10.1021/ja00092a048},
  shorttitle   = {Conformational flexibility of \emph{O}-phosphorylcholine and \emph{o}-phosphorylethanolamine}
}

@techreport{WorldMeteorologicalOrganizationWMO.WorldMeteorologicalOrganizationWMO.2025.StateGlobalClimate,

  author       = {{World Meteorological Organization (WMO)}},
  title        = {State of the Global Climate 2024},
  institution  = {WMO},
  address      = {Geneva},
  number       = {WMO-No. 1368},
  year         = {2025}
}

@article{Zhao.Hussein.2025.LongtermPredictionClimate,

  author       = {Zhao, Jiangyue and Salthammer, Tunga and Schieweck, Alexandra and Uhde, Erik and Hussein, Tareq},
  title        = {Long-term prediction of climate change impacts on indoor particle pollution: Case study of a residential building in Germany},
  journal      = {Environ. Sci.: Processes Impacts},
  year         = {2025},
  month        = jun,
  volume       = {27},
  number       = {6},
  pages        = {1688--1703},
  doi          = {10.1039/D5EM00249A}
}

@article{Zhao.Schieweck.2024.LongtermPredictionEffects,

  author       = {Zhao, Jiangyue and Uhde, Erik and Salthammer, Tunga and Antretter, Florian and Shaw, David and Carslaw, Nicola and Schieweck, Alexandra},
  title        = {Long-term prediction of the effects of climate change on indoor climate and air quality},
  journal      = {Environ. Res.},
  year         = {2024},
  month        = feb,
  volume       = {243},
  pages        = {117804},
  doi          = {10.1016/j.envres.2024.117804}
}

@article{anderson_heterogeneity_1988,

  author       = {Anderson, B. D. and Higuchi, W. I. and Raykar, P. V.},
  title        = {Heterogeneity effects on permeability--partition coefficient relationships in human stratum corneum},
  journal      = {Pharm. Res.},
  year         = {1988},
  month        = sep,
  volume       = {5},
  number       = {9},
  pages        = {566--573},
  doi          = {10.1023/A:1015989929342}
}

@article{anderson_solute_1989,

  author       = {Anderson, Bradley D. and Raykar, Prakash V.},
  title        = {Solute structure--permeability relationships in human stratum corneum},
  journal      = {J. Invest. Dermatol.},
  year         = {1989},
  month        = aug,
  volume       = {93},
  number       = {2},
  pages        = {280--286},
  doi          = {10.1111/1523-1747.ep12277592}
}

@article{bussi_canonical_2007,

  author       = {Bussi, Giovanni and Donadio, Davide and Parrinello, Michele},
  title        = {Canonical sampling through velocity rescaling},
  journal      = {J. Chem. Phys.},
  year         = {2007},
  month        = jan,
  volume       = {126},
  number       = {1},
  pages        = {014101},
  doi          = {10.1063/1.2408420}
}

@article{gaalswyk_generalized_2016,

  author       = {Gaalswyk, Kari and Awoonor-Williams, Ernest and Rowley, Christopher N.},
  title        = {Generalized Langevin methods for calculating transmembrane diffusivity},
  journal      = {J. Chem. Theory Comput.},
  year         = {2016},
  volume       = {12},
  number       = {11},
  pages        = {5609--5619},
  doi          = {10.1021/acs.jctc.6b00747},
  journaltitle = {Journal of Chemical Theory and Computation},
  shortjournal = {J. Chem. Theory Comput.},
  date         = {2016-11-08}
}

@article{gupta_molecular_2016,

  author       = {Gupta, Rakesh and Sridhar, D. B. and Rai, Beena},
  title        = {Molecular dynamics simulation study of permeation of molecules through skin lipid bilayer},
  journal      = {J. Phys. Chem. B},
  year         = {2016},
  month        = sep,
  volume       = {120},
  number       = {34},
  pages        = {8987--8996},
  doi          = {10.1021/acs.jpcb.6b05451}
}

@article{hummer_position-dependent_2005,

  author       = {Hummer, Gerhard},
  title        = {Position-dependent diffusion coefficients and free energies from Bayesian analysis of equilibrium and replica molecular dynamics simulations},
  journal      = {New J. Phys.},
  year         = {2005},
  month        = jan,
  volume       = {7},
  number       = {1},
  pages        = {34},
  doi          = {10.1088/1367-2630/7/1/034}
}

@article{kalia_homogeneous_1996,

  author       = {Kalia, Y. N. and Pirot, F. and Guy, R. H.},
  title        = {Homogeneous transport in a heterogeneous membrane: Water diffusion across human stratum corneum in vivo},
  journal      = {Biophys. J.},
  year         = {1996},
  month        = nov,
  volume       = {71},
  number       = {5},
  pages        = {2692--2700},
  doi          = {10.1016/S0006-3495(96)79460-2}
}

@article{lakey_kinetic_2021,

  author       = {Lakey, Pascale S. J. and Eichler, Clara M. A. and Wang, Chunyi and Little, John C. and Shiraiwa, Manabu},
  title        = {Kinetic multi-layer model of film formation, growth, and chemistry (KM-FILM): Boundary layer processes, multi-layer adsorption, bulk diffusion, and heterogeneous reactions},
  journal      = {Indoor Air},
  year         = {2021},
  month        = nov,
  volume       = {31},
  number       = {6},
  pages        = {2070--2083},
  doi          = {10.1111/ina.12854},
  shorttitle   = {Kinetic multi-layer model of film formation, growth, and chemistry (KM-FILM)}
}

@article{lian_evaluation_2008,

  author       = {Lian, Guoping and Chen, Longjian and Han, Lujia},
  title        = {An evaluation of mathematical models for predicting skin permeability},
  journal      = {J. Pharm. Sci.},
  year         = {2008},
  month        = jan,
  volume       = {97},
  number       = {1},
  pages        = {584--598},
  doi          = {10.1002/jps.21074}
}

@article{phillips_scalable_2005,

  author       = {Phillips, James C. and Braun, Rosemary and Wang, Wei and Gumbart, James and Tajkhorshid, Emad and Villa, Elizabeth and Chipot, Christophe and Skeel, Robert D. and Kal{\'e}, Laxmikant and Schulten, Klaus},
  title        = {Scalable molecular dynamics with NAMD},
  journal      = {J. Comput. Chem.},
  year         = {2005},
  volume       = {26},
  number       = {16},
  pages        = {1781--1802},
  doi          = {10.1002/jcc.20289}
}

@article{phillips_scalable_2020,

  author       = {Phillips, James C. and Hardy, David J. and Maia, Julio D. C. and Stone, John E. and Ribeiro, Jo{\~a}o V. and Bernardi, Rafael C. and Buch, Ronak and Fiorin, Giacomo and H{\'e}nin, J{\'e}r{\^o}me and Jiang, Wei and McGreevy, Ryan and Melo, Marcelo C. R. and Radak, Brian K. and Skeel, Robert D. and Singharoy, Abhishek and Wang, Yi and Roux, Beno{\^\i}t and Aksimentiev, Aleksei and Luthey-Schulten, Zaida and Kal{\'e}, Laxmikant V. and Schulten, Klaus and Chipot, Christophe and Tajkhorshid, Emad},
  title        = {Scalable molecular dynamics on CPU and GPU architectures with NAMD},
  journal      = {J. Chem. Phys.},
  year         = {2020},
  month        = jul,
  volume       = {153},
  number       = {4},
  pages        = {044130},
  doi          = {10.1063/5.0014475}
}

@article{thomas_insights_2025,

  author       = {Thomas, Rinto and Prabhakar, Praveen Ranganath and Tobias, Douglas J. and von Domaros, Michael},
  title        = {Insights into dermal permeation of skin oil oxidation products from enhanced sampling molecular dynamics simulation},
  journal      = {J. Phys. Chem. B},
  year         = {2025},
  volume       = {129},
  number       = {6},
  pages        = {1784--1794},
  doi          = {10.1021/acs.jpcb.4c08090}
}

@Article{wang_models_2018,

  author       = {Wang, Eric and Klauda, Jeffery B.},
  title        = {Models for the stratum corneum lipid matrix: Effects of ceramide concentration, ceramide hydroxylation, and free fatty acid protonation},
  journal      = {J. Phys. Chem. B},
  year         = {2018},
  month        = dec,
  volume       = {122},
  number       = {50},
  pages        = {11996--12008},
  issn         = {1520-6106},
  doi          = {10.1021/acs.jpcb.8b06188}
}

@article{wilschut_estimating_1995,

  author       = {Wilschut, Annette and ten Berge, Wil F. and Robinson, Peter J. and McKone, Thomas E.},
  title        = {Estimating skin permeation: The validation of five mathematical skin permeation models},
  journal      = {Chemosphere},
  year         = {1995},
  month        = apr,
  volume       = {30},
  number       = {7},
  pages        = {1275--1296},
  doi          = {10.1016/0045-6535(95)00023-2}
}

@article{Comer.Gonzalez-Nilo.2013.CalculatingPositionDependentDiffusivity,
  author       = {Comer, Jeffrey and Chipot, Christophe and González-Nilo, Fernando D.},
  title        = {Calculating position-dependent diffusivity in biased molecular dynamics simulations},
  journal      = {J. Chem. Theory Comput.},
  year         = {2013},
  month        = feb,
  volume       = {9},
  number       = {2},
  pages        = {876--882},
  doi          = {10.1021/ct300706c}
}

@article{Ghysels.Hummer.2017.PositionDependentDiffusionTensors,
  author       = {Ghysels, An and Venable, Richard M. and Pastor, Richard W. and Hummer, Gerhard},
  title        = {Position-dependent diffusion tensors in anisotropic media from simulation: Oxygen transport in and through membranes},
  journal      = {J. Chem. Theory Comput.},
  year         = {2017},
  month        = jun,
  volume       = {13},
  number       = {6},
  pages        = {2962--2976},
  doi          = {10.1021/acs.jctc.7b00175}
}

@article{Rosta.Hummer.2015.FreeEnergiesDynamic,
  author       = {Rosta, Edina and Hummer, Gerhard},
  title        = {Free energies from dynamic weighted histogram analysis using unbiased Markov state models},
  journal      = {J. Chem. Theory Comput.},
  year         = {2015},
  month        = jan,
  volume       = {11},
  number       = {1},
  pages        = {276--285},
  doi          = {10.1021/ct500719p}
}

@article{Sicard.Rosta.2021.PositionDependentDiffusionBiased,
  author       = {Sicard, François and Koskin, Vladimir and Annibale, Alessia and Rosta, Edina},
  title        = {Position-dependent diffusion from biased simulations and Markov state model analysis},
  journal      = {J. Chem. Theory Comput.},
  year         = {2021},
  month        = apr,
  volume       = {17},
  number       = {4},
  pages        = {2022--2033},
  doi          = {10.1021/acs.jctc.0c01212}
}

@article{piasentin_evaluation_2021,
  author       = {Piasentin, Nicola and Lian, Guoping and Cai, Qiong},
  title        = {Evaluation of constrained and restrained molecular dynamics simulation methods for predicting skin lipid permeability},
  journal      = {ACS Omega},
  year         = {2021},
  month        = dec,
  volume       = {6},
  number       = {51},
  pages        = {35363--35374},
  doi          = {10.1021/acsomega.1c04684}
}

@article{kramer_membrane_2020,
  author       = {Krämer, Andreas and Ghysels, An and Wang, Eric and Venable, Richard M. and Klauda, Jeffery B. and Brooks, Bernard R. and Pastor, Richard W.},
  title        = {Membrane permeability of small molecules from unbiased molecular dynamics simulations},
  journal      = {J. Chem. Phys.},
  year         = {2020},
  month        = sep,
  volume       = {153},
  number       = {12},
  pages        = {124107},
  doi          = {10.1063/5.0013429}
}

@article{das_water_2009,
  author       = {Das, Chinmay and Olmsted, Peter D. and Noro, Massimo G.},
  title        = {Water permeation through stratum corneum lipid bilayers from atomistic simulations},
  journal      = {Soft Matter},
  year         = {2009},
  month        = nov,
  volume       = {5},
  number       = {22},
  pages        = {4549--4555},
  doi          = {10.1039/B911257J}
}

@article{Mitragotri.Roberts.2011,
  author       = {Mitragotri, Samir and Anissimov, Yuri G. and Bunge, Annette L. and Frasch, H. Frederick and Guy, Richard H. and Hadgraft, Jonathan and Kasting, Gerald B. and Lane, Majella E. and Roberts, Michael S.},
  title        = {Mathematical models of skin permeability: An overview},
  journal      = {Int. J. Pharm.},
  year         = {2011},
  volume       = {418},
  number       = {1},
  pages        = {115--129},
  doi          = {10.1016/j.ijpharm.2011.02.023}
}

@article{Frasch.Barbero.2013,
  author       = {Frasch, H. Frederick and Barbero, Ana M.},
  title        = {Application of numerical methods for diffusion-based modeling of skin permeation},
  journal      = {Adv. Drug Delivery Rev.},
  year         = {2013},
  volume       = {65},
  number       = {2},
  pages        = {208--220},
  doi          = {10.1016/j.addr.2012.01.001}
}

@article{Wang.Naegel.2023,
  author       = {Wang, Junxi and Nitsche, Johannes M. and Kasting, Gerald B. and Wittum, Gabriel and Nägel, Arne},
  title        = {Transdermal and lateral effective diffusivities for drug transport in stratum corneum from a microscopic anisotropic diffusion model},
  journal      = {Eur. J. Pharm. Biopharm.},
  year         = {2023},
  volume       = {188},
  pages        = {271--286},
  doi          = {10.1016/j.ejpb.2023.01.025}
}

@article{Reuter.Lunter.2025.PresenceDifferentCeramide,
  author       = {Reuter, Moritz and Joseph, Edwin and Lian, Guoping and Lunter, Dominique J.},
  title        = {Presence of different ceramide species modulates barrier function and structure of stratum corneum lipid membranes: Insights from molecular dynamics simulations},
  journal      = {Mol. Pharm.},
  year         = {2025},
  month        = jul,
  volume       = {22},
  number       = {7},
  pages        = {4280--4292},
  doi          = {10.1021/acs.molpharmaceut.5b00457}
}

@article{Michaels.Shaw.1975.DrugPermeationHuman,
  author  = {Michaels, A. S. and Chandrasekaran, S. K. and Shaw, J. E.},
  title   = {Drug permeation through human skin: Theory and in vitro experimental measurement},
  journal = {AIChE J.},
  year    = {1975},
  volume  = {21},
  number  = {5},
  pages   = {985--996},
  doi     = {10/bj5fd9}
}

@article{Elias.Elias.1983.EpidermalLipidsBarrier,
  author  = {Elias, Peter M.},
  title   = {Epidermal lipids, barrier function, and desquamation},
  journal = {J. Invest. Dermatol.},
  year    = {1983},
  volume  = {80},
  number  = {1},
  pages   = {S44--S49},
  doi     = {10/dxbtmb}
}

@article{Schmitt.Neubert.2020.StateArtStratum,
  author  = {Schmitt, Thomas and Neubert, Reinhard H. H.},
  title   = {State of the art in stratum corneum research. Part {II}: Hypothetical stratum corneum lipid matrix models},
  journal = {Skin Pharmacol. Physiol.},
  year    = {2020},
  volume  = {33},
  number  = {4},
  pages   = {213--230},
  doi     = {10/grcws5}
}

@article{Wang.Nagel.2023.TransdermalLateralEffective,
  author  = {Wang, Junxi and Nitsche, Johannes M. and Kasting, Gerald B. and Wittum, Gabriel and N{\"a}gel, Arne},
  title   = {Transdermal and lateral effective diffusivities for drug transport in stratum corneum from a microscopic anisotropic diffusion model},
  journal = {Eur. J. Pharm. Biopharm.},
  year    = {2023},
  volume  = {188},
  pages   = {271--286},
  doi     = {10/g8zqc5}
}

@article{Wang.Nitsche.2007.MultiphaseMicroscopicDiffusion,
  author  = {Wang, Tsuo-Feng and Kasting, Gerald B. and Nitsche, Johannes M.},
  title   = {A multiphase microscopic diffusion model for stratum corneum permeability. {II}. Estimation of physicochemical parameters, and application to a large permeability database},
  journal = {J. Pharm. Sci.},
  year    = {2007},
  volume  = {96},
  number  = {11},
  pages   = {3024--3051},
  doi     = {10/fbh7b5}
}

@article{Bouwstra.Gooris.2023.SkinBarrierExtraordinary,
  author  = {Bouwstra, Joke A. and N\u{a}d\u{a}ban, Andreea and Bras, Wim and McCabe, Clare and Bunge, Annette and Gooris, Gerrit S.},
  title   = {The skin barrier: An extraordinary interface with an exceptional lipid organization},
  journal = {Prog. Lipid Res.},
  year    = {2023},
  volume  = {92},
  pages   = {101252},
  doi     = {10/gt72qw}
}

@article{Sparr.Topgaard.2023.StratumCorneumBarrier,
  author  = {Sparr, Emma and Bj{\"o}rklund, Sebastian and Pham, Q. Dat and Mojumdar, Enamul H. and Stenqvist, B. and Gunnarsson, M. and Topgaard, D.},
  title   = {The stratum corneum barrier -- from molecular scale to macroscopic properties},
  journal = {Curr. Opin. Colloid Interface Sci.},
  year    = {2023},
  volume  = {67},
  pages   = {101725},
  doi     = {10/gsvb2j}
}

@article{reuter_presence_2025,
  author  = {Reuter, Moritz and Joseph, Edwin and Lian, Guoping and Lunter, Dominique J.},
  title   = {Presence of different ceramide species modulates barrier function and structure of stratum corneum lipid membranes: Insights from molecular dynamics simulations},
  journal = {Mol. Pharm.},
  year    = {2025},
  volume  = {22},
  number  = {7},
  pages   = {4280--4292},
  doi     = {10.1021/acs.molpharmaceut.5c00580}
}

@article{katritzky_skin_2006,
  author  = {Katritzky, Alan R. and Dobchev, Dimitar A. and Fara, Dan C. and H{\"u}r, Evrim and T{\"a}mm, Kaido and Kurunczi, Ludovic and Karelson, Mati and Varnek, Alexandre and {Solov'ev}, Vitaly P.},
  title   = {Skin permeation rate as a function of chemical structure},
  journal = {J. Med. Chem.},
  year    = {2006},
  volume  = {49},
  number  = {11},
  pages   = {3305--3314},
  doi     = {10.1021/jm051031d}
}

@article{chen_modeling_2010,
  author  = {Chen, Longjian and Lian, Guoping and Han, Lujia},
  title   = {Modeling transdermal permeation. Part {I}. Predicting skin permeability of both hydrophobic and hydrophilic solutes},
  journal = {AIChE J.},
  year    = {2010},
  volume  = {56},
  number  = {5},
  pages   = {1136--1146},
  doi     = {10.1002/aic.12048}
}

@article{FujimotoNagaiEtAl2021,
  author  = {Fujimoto, Kazushi and Nagai, Tetsuro and Yamaguchi, Tsuyoshi},
  title   = {Momentum removal to obtain the position-dependent diffusion constant in constrained molecular dynamics simulation},
  journal = {J. Comput. Chem.},
  year    = {2021},
  volume  = {42},
  number  = {30},
  pages   = {2136--2144},
  doi     = {10.1002/jcc.26742}
}

@article{jamali_finite-size_2018,
  author       = {Jamali, Seyed Hossein and Wolff, Ludger and Becker, Tim M. and Bardow, André and Vlugt, Thijs J. H. and Moultos, Othonas A.},
  title        = {Finite-size effects of binary mutual diffusion coefficients from molecular dynamics},
  journal      = {J. Chem. Theory Comput.},
  year         = {2018},
  month        = may,
  volume       = {14},
  number       = {5},
  pages        = {2667--2677},
  doi          = {10.1021/acs.jctc.8b00170}
}

@article{martinez_p_2009,
  author       = {Mart{\'i}nez, L. and Andrade, R. and Birgin, E. G. and Mart{\'i}nez, J. M.},
  title        = {PACKMOL: A package for building initial configurations for molecular dynamics simulations},
  journal      = {J. Comput. Chem.},
  year         = {2009},
  month        = oct,
  volume       = {30},
  number       = {13},
  pages        = {2157--2164},
  doi          = {10.1002/jcc.21224}
}

@article{Yeh.Hummer.2004.SystemSizeDependenceDiffusion,
  author       = {Yeh, In-Chul and Hummer, Gerhard},
  title        = {System-Size Dependence of Diffusion Coefficients and Viscosities from Molecular Dynamics Simulations with Periodic Boundary Conditions},
  journal      = {J. Phys. Chem. B},
  year         = {2004},
  month        = oct,
  volume       = {108},
  number       = {40},
  pages        = {15873--15879},
  publisher    = {American Chemical Society},
  doi          = {10.1021/jp0477147}
}

\end{document}


\maketitle

\section{Impact of the Harmonic Force Constant on Diffusivity Profiles}

A harmonic potential is required to restrain the solute at selected positions along the permeation path and to enable the VACF- and PACF-based analyses described in the main text. 
We employed three different force constants, \SIlist{2.5;30.0;50.0}{\kcal\per\mol\per\angstrom\squared}, which, in principle, should yield identical results. 
However, the strength of the restraint potential directly affects the behavior of both correlation functions, influencing their decay rates and corresponding Laplace spectra. 
Different force constants may therefore introduce numerical artifacts by altering the signal-to-noise ratio. 
A detailed discussion of these effects was provided by Rowley \textit{et~al.}\cite{gaalswyk_generalized_2016} 

\begin{center}
    \includegraphics[width=\textwidth]{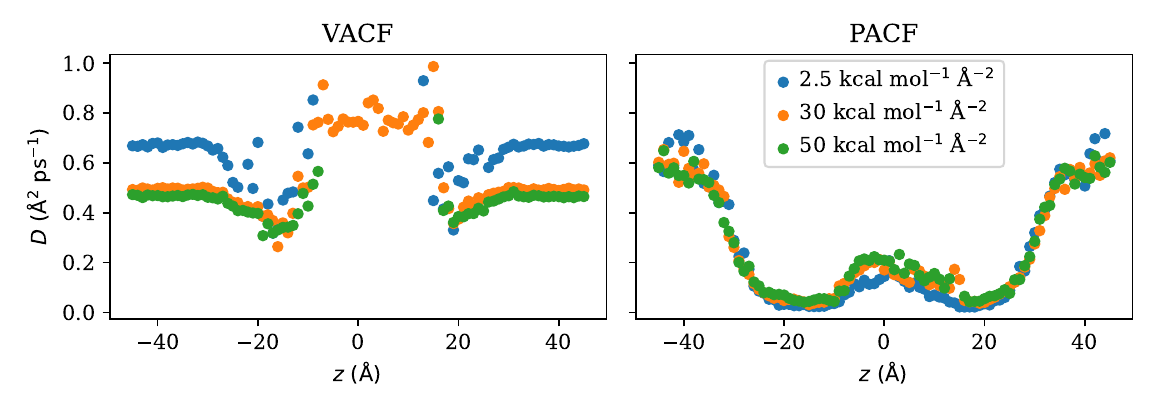}
    \captionof{figure}{
        Diffusivity profiles of water obtained from the PACF and VACF analyses 
        in the SC membrane at three different harmonic force constants.
    }
    \label{fig:force_constants}
\end{center}

Our results show similar trends, with modest variations in PACF-derived diffusivities across force constants but substantially larger variations in the VACF-derived values (Figure~\ref{fig:force_constants}). 
In addition, some VACF-derived diffusivity profiles exhibit clear artifacts, likely stemming from limitations in the automated extrapolation procedure employed. 
We chose a force constant of \SI{30}{\kcal\per\mol\per\angstrom\squared} for the results presented in the main text, as this value produced the most stable and physically consistent profiles. 
Although more accurate VACF-based diffusivity profiles could, in principle, be obtained by carefully analyzing individual VACFs and their Laplace transforms and by manually tuning the extrapolation procedure, such an approach is laborious, error-prone, and potentially biased. These observations underscore the limited practicality of the VACF method.

\section{Bulk Diffusivities}

To validate the diffusivity profiles presented in the main text, we calculated the diffusivities of all solutes in bulk water and compared them with those obtained at the center of the aqueous phase in our membrane system.

\subsection{Method}

A single molecule of each solute and \num{2832} water molecules were placed in a rectangular simulation cell with dimensions of \SI{57.15}{\angstrom} \(\times\) \SI{57.15}{\angstrom} \(\times\) \SI{26}{\angstrom}
, resulting in a density of \qty{1.00}{\gram\per\centi\meter\cubed} for all systems, corresponding to the experimental density of water at a temperature of $T = \qty{305.15}{\kelvin}$ and a pressure of $P = {\qty{1}{\bar}}$. 
The chosen box size approximately matches the size of the bulk water region in our membrane system. 
The systems were constructed using Packmol.\cite{martinez_p_2009}

Unless noted otherwise, the simulation parameters were consistent with those used in our previous work.\cite{thomas_insights_2025}
Each system was first energy-minimized for \num{50000} steps, followed by a short equilibration run of \qty{1}{\nano\second} in the NVT ensemble and a subsequent \qty{20}{\nano\second} simulation in the NPT ensemble. 
The system was then equilibrated for an additional \qty{20}{\nano\second} under NVT conditions prior to production runs. 
A total of \num{14} independent production trajectories of \qty{1}{\nano\second} each were performed, during which the center-of-mass coordinates of the solutes were recorded every \qty{2}{\femto\second}. Each \qty{1}{\nano\second} trajectory was subdivided into four \qty{250}{\pico\second} segments, yielding \num{56} segments per solute.

In homogeneous bulk systems, the Einstein relation connects the mean-squared displacement~(MSD) to the diffusion coefficient $D$ in the limit of long lag times $t$,
\begin{equation}
    \lim_{t \to \infty} \left< \left| \vec{r}(t) - \vec{r}(0) \right|^{2} \right> = 2 d D t,
    \label{eq:einstein}
\end{equation}
where $\vec{r}$ denotes the center-of-mass position of the particle of interest, and $d$ is the dimensionality of the system.

\subsection{Results and Discussion}

Diffusion is inherently a size-dependent phenomenon,\cite{Yeh.Hummer.2004.SystemSizeDependenceDiffusion,jamali_finite-size_2018} which is particularly important in the finite aqueous region of our membrane system. 
We therefore evaluated bulk diffusivities in a rectangular simulation cell of comparable size. 
In addition, we analyzed both one-dimensional ($d = 1$) diffusion along the $z$-axis of the simulation cell—most closely resembling the setup of the membrane system—and three-dimensional ($d = 3$) diffusion, which is the more common approach for studying bulk diffusion as it provides improved statistical sampling.

\begin{center}
    \includegraphics[width=\textwidth]{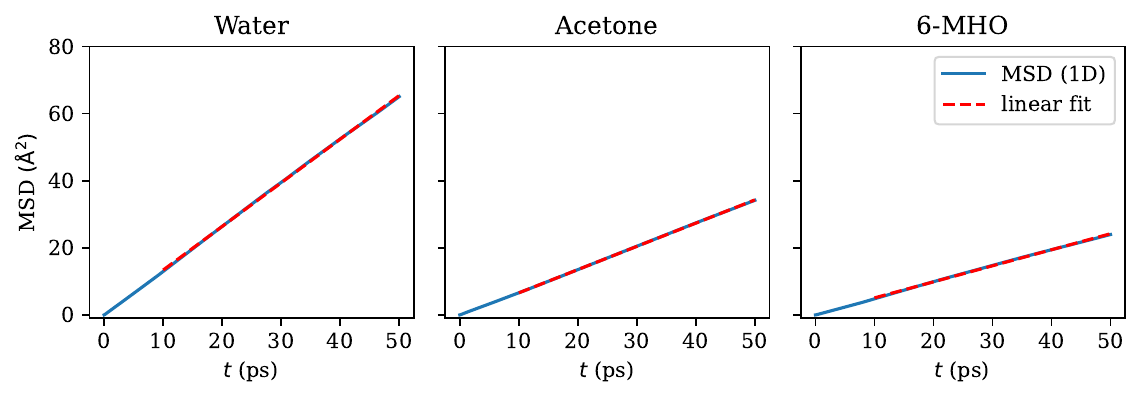}
    \captionof{figure}{
        One-dimensional mean-squared displacements (MSDs) along the $z$-axis of the simulation cell for all solutes in bulk water.
    }
    \label{fig:msd_1d}
\end{center}

\begin{center}
    \includegraphics[width=\textwidth]{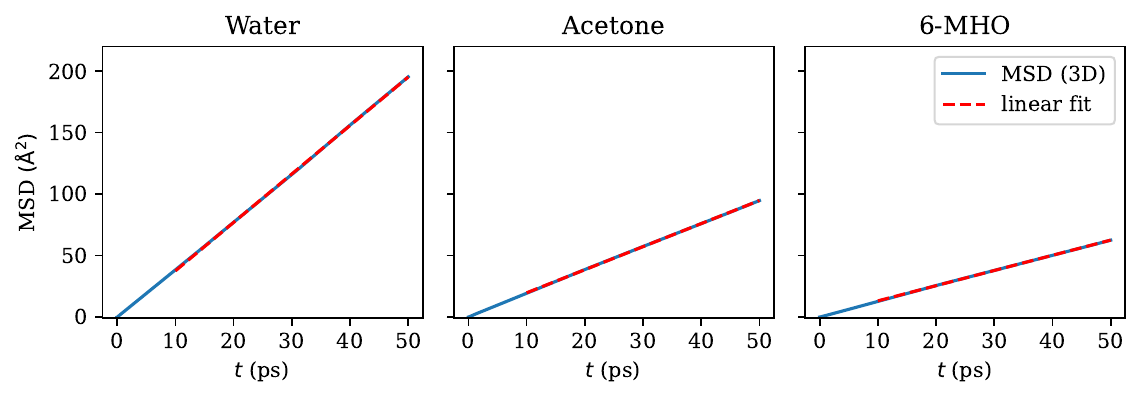}
    \captionof{figure}{
        Three-dimensional mean-squared displacements (MSDs) for all solutes in bulk water.
    }
    \label{fig:msd_3d}
\end{center}

\begin{center}
\captionof{table}{
    Bulk diffusion coefficients (in $\qty{}{\angstrom^2\per\pico\second}$) for all solutes in bulk water, 
    computed from the Einstein relation (eq.~\ref{eq:einstein}). 
    Uncertainties represent the standard error of the mean across independent simulations 
    and correspond to a 95\% confidence interval.
}
\label{tab:diffusion_constants}
\begin{tabular*}{\columnwidth}{@{\extracolsep{\fill}}lrr}
    \toprule
    Solute & 1D & 3D \\ 
    \midrule
    Water   & $0.650 \pm 0.109$ & $0.657 \pm 0.077$ \\ 
    Acetone & $0.346 \pm 0.055$ & $0.313 \pm 0.033$ \\
    6-MHO   & $0.239 \pm 0.048$ & $0.206 \pm 0.021$ \\
    \bottomrule
\end{tabular*}
\end{center}

Mean-squared displacements~(MSDs) for the one-dimensional case are shown in Figure~\ref{fig:msd_1d}, and those for the three-dimensional case are presented in Figure~\ref{fig:msd_3d}. 
Bulk diffusivities, obtained from linear fits to the Einstein relation (eq.~\ref{eq:einstein}), are summarized in Table~\ref{tab:diffusion_constants}. 
For all fits, only lag times between \qty{10}{\pico\second} and \qty{50}{\pico\second} were considered.

Among the three solutes, the bulk diffusion coefficient is highest for water, followed by acetone and 6-MHO, consistent with the decrease in hydrodynamic radius across this series.

\section{Propagator Analysis}

In Figure~\ref{fig:all_propagators}, we present the full set of propagators, $p(z, t \,|\, z_0 = 0, t_0 = 0)$, computed in the critical membrane-center region from both diffusion models for selected lag times~$t$. 
The propagators were obtained using diffusivity profiles derived from the velocity autocorrelation function~(VACF) and position autocorrelation function~(PACF), in combination with the umbrella sampling~(US) free-energy profiles shown in the main text. 
Between $t = \SI{50}{\pico\second}$ and $t = \SI{500}{\pico\second}$, the results qualitatively match the representative example at $t = \SI{100}{\pico\second}$ shown in the main text: neither the VACF- nor the PACF-based model reproduces the simulation data quantitatively. 
The VACF results consistently overestimate the diffusive spread, whereas the PACF results underestimate it.

\begin{center}
    \includegraphics[width=\textwidth]{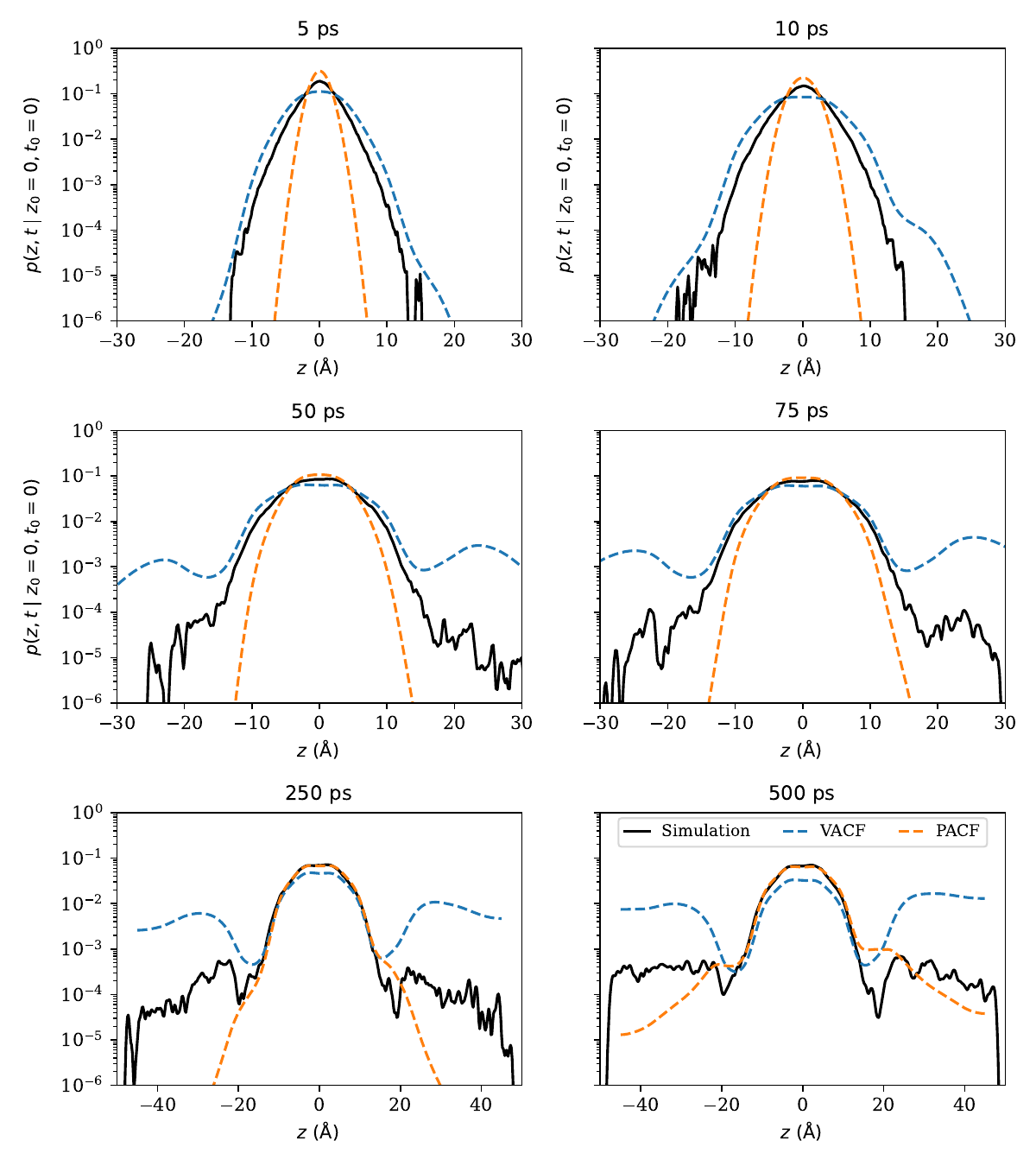}
    \captionof{figure}{
        Propagators, $p(z, t \,|\, z_0, t_0)$, at different lag times ranging from 
        $t = \SI{5}{\pico\second}$ to $t = \SI{500}{\pico\second}$, obtained from simulations and diffusion models based on VACF- and PACF-derived diffusivity profiles combined with US free-energy profiles.
    }
    \label{fig:all_propagators}
\end{center}

\section{Permeabilities}

The estimation of membrane permeabilities using the inhomogeneous solubility–diffusion (ISD) model requires defining the membrane extent, $h$. 
Because of the soft and fluctuating nature of lipid membranes, establishing precise boundaries is challenging. 
It is common practice to define the interfacial region based on the density profile, which is shown in Figure~\ref{fig:density_profile} for our SC model. 
However, there is no well-established criterion for selecting cutoff positions. 
In the absence of a standard definition, we considered three plausible choices for $h/2$: 
(\textit{i}) the position where the water density begins to decrease from its bulk value ($h_1/2 = \qty{32.0}{\angstrom}$); 
(\textit{ii}) the position where the water density is reduced to half of its bulk value ($h_2/2 = \qty{25.5}{\angstrom}$); and 
(\textit{iii}) the position where the water density approaches zero ($h_3/2 = \qty{17.0}{\angstrom}$).

\begin{center}
    \includegraphics[]{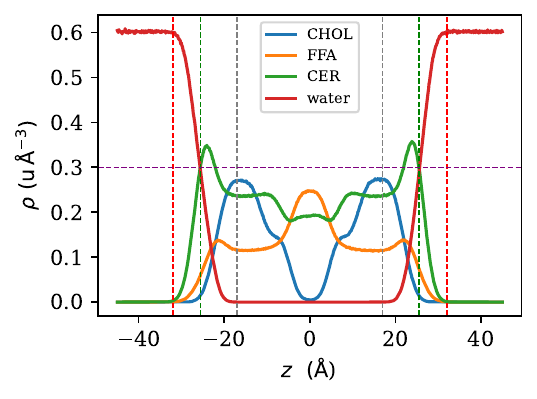}
    \captionof{figure}{
        Component-wise density profile of the stratum corneum~(SC) membrane 
        (cholesterol, CHOL, blue; lignoceric acid, FFA, orange; ceramide~NS, CER, green; and water, red). 
        The red, green, and gray vertical dashed lines indicate the different integration limits tested at 
        $h_{1}/2 = \pm \qty{32.0}{\angstrom}$, 
        $h_{2}/2 = \pm \qty{25.5}{\angstrom}$, 
        and $h_{3}/2 = \pm \qty{17.0}{\angstrom}$, respectively. 
        The purple horizontal dashed line corresponds to half of the bulk water density.
    }
    \label{fig:density_profile}
\end{center}

Below we present the local resistance profiles, $R(z)$, for the combinations of free-energy and diffusivity methods not shown in the main text: US/PACF (Figure~\ref{fig:resistance_us_pacf}), WTM/VACF (Figure~\ref{fig:resistance_wtm_vacf}), and WTM/PACF (Figure~\ref{fig:resistance_wtm_pacf}). In all cases, the conclusion from the main text holds: the additional contribution to the area under the curve—and thus to the inverse permeability, $P^{-1}$—is negligible when changing from the smallest to the largest membrane extent considered. Any of the reasonable cutoff definitions therefore has little to no impact on the computed permeability.

\begin{center}
    \includegraphics[width=\columnwidth]{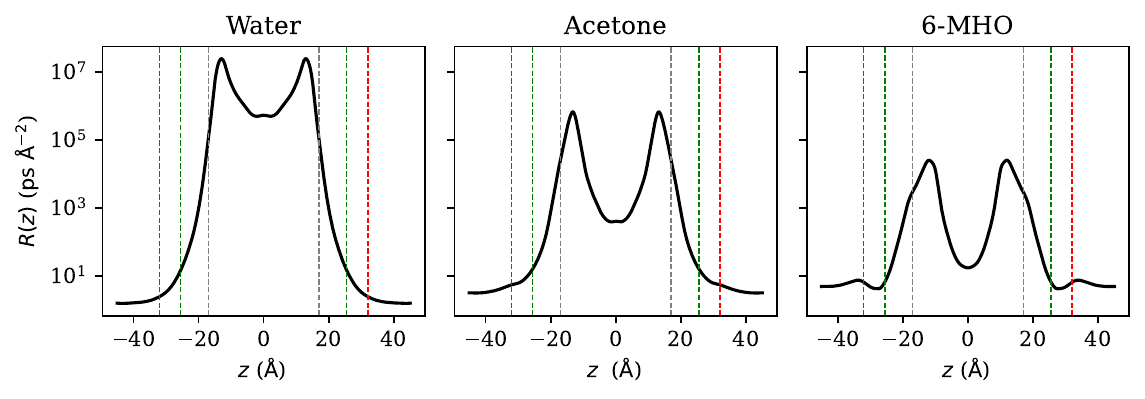}
    \captionof{figure}{
        Local resistance, $R(z)$, estimated using the combination of umbrella sampling~(US) free-energy profiles 
        and PACF-derived diffusivities. 
        Dashed vertical lines correspond to the different membrane extents defined in Figure~\ref{fig:density_profile}.
    }
    \label{fig:resistance_us_pacf}
\end{center}

\begin{center}
    \includegraphics[width=\columnwidth]{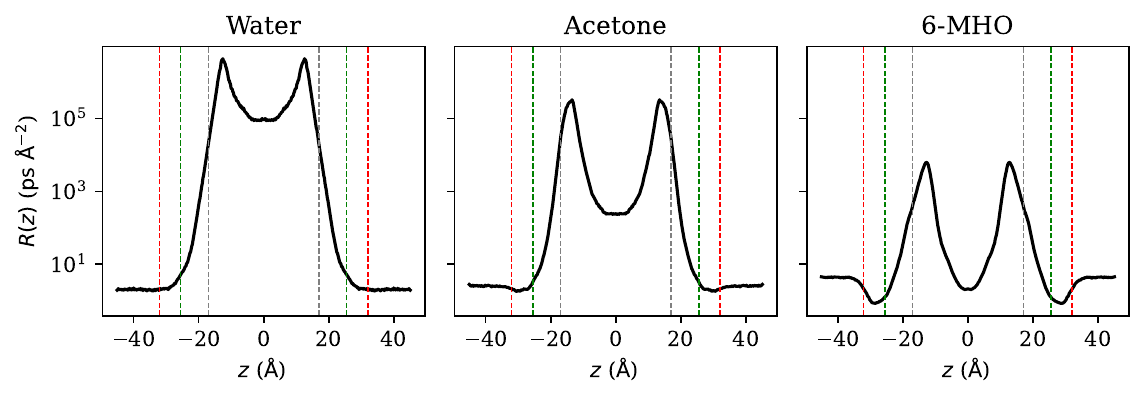}
    \captionof{figure}{
        Local resistance, $R(z)$, estimated using the combination of well-tempered metadynamics~(WTM) free-energy profiles 
        and VACF-derived diffusivities. 
        Dashed vertical lines correspond to the different membrane extents defined in Figure~\ref{fig:density_profile}.
    }
    \label{fig:resistance_wtm_vacf}
\end{center}

\noindent
\begin{minipage}{\columnwidth}
    \centering
    \includegraphics[width=\columnwidth]{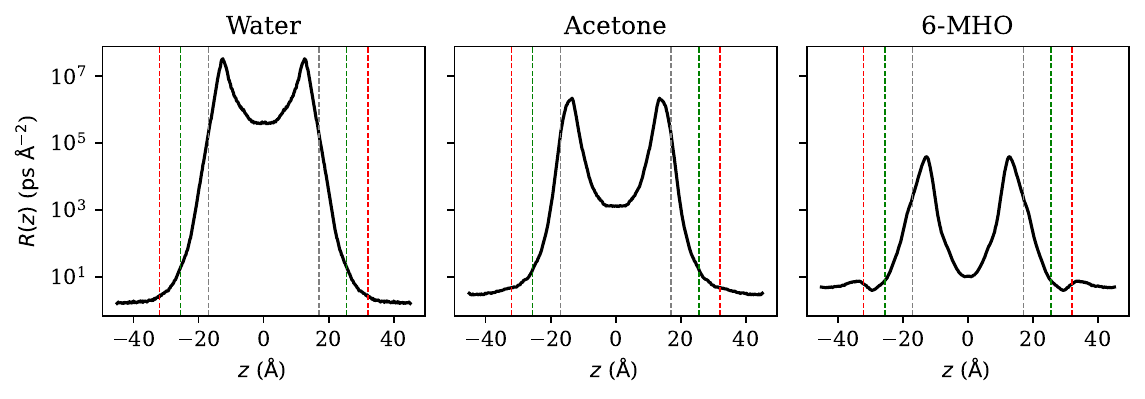}
    \captionof{figure}{
        Local resistance, $R(z)$, estimated using the combination of well-tempered metadynamics~(WTM) 
        free-energy profiles and PACF-derived diffusivities. 
        Dashed vertical lines correspond to the different membrane extents defined in Figure~\ref{fig:density_profile}.
    }
    \label{fig:resistance_wtm_pacf}
\end{minipage}
\vspace{1em}

Finally, we present in Table~\ref{tab:permeability} the numerical values of the membrane permeabilities that are plotted graphically in the main text.

\begin{center}
\captionof{table}{
    Membrane permeabilities (in \si{\centi\meter\per\second}) calculated using position autocorrelation function (PACF) and velocity autocorrelation function (VACF) methods. Free-energy profiles were obtained from umbrella sampling (US) and well-tempered metadynamics (WTM). The corresponding bar plots are shown in the main text.
}
\label{tab:permeability}
\begin{tabular*}{\columnwidth}{@{\extracolsep{\fill}}lcccc}
    \toprule
     & \multicolumn{2}{c}{PACF} & \multicolumn{2}{c}{VACF} \\
    \cmidrule(lr){2-3} \cmidrule(lr){4-5}
    Solute & US & WTM & US & WTM \\
    \midrule    
    Water   & $5.65 \times 10^{-5}$ & $5.48 \times 10^{-5}$ & $4.15 \times 10^{-4}$ & $4.00 \times 10^{-4}$ \\ 
    Acetone & $2.60 \times 10^{-3}$ & $7.48 \times 10^{-4}$ & $1.75 \times 10^{-2}$ & $5.04 \times 10^{-3}$ \\
    6-MHO   & $4.55 \times 10^{-2}$ & $4.18 \times 10^{-2}$ & $2.82 \times 10^{-1}$ & $2.60 \times 10^{-1}$ \\
    \bottomrule
\end{tabular*}
\end{center}

\bibliography{si}